\documentclass[graybox]{svmult}

\usepackage{type1cm} % activate if the above 3 fonts are not available on your system
\usepackage{makeidx} % allows index generation
\usepackage{graphicx} % standard LaTeX graphics tool when including figure files
\usepackage{multicol} % used for the two-column index
\usepackage[bottom]{footmisc} % places footnotes at page bottom

\usepackage{newtxtext}
\usepackage[varvw]{newtxmath} % selects Times Roman as basic font

\usepackage[colorlinks=true,urlcolor=blue]{hyperref}
\usepackage{siunitx}

\DeclareSIUnit{\year}{yr}
\DeclareSIUnit{\ME}{M_{\oplus}}
\DeclareSIUnit{\RM}{R_{M}}
\DeclareSIUnit{\MM}{M_{M}} % if this is changed to not use \mercury, remove the package wasysym

\usepackage[normalem]{ulem}

\makeindex % used for the subject index please use the style svind.ist with your makeindex program

\begin{document}
\date{\today}
\title*{Recent advances in modelling of global-scale collisions using smoothed particle hydrodynamics}

\author{Christian Reinhardt\orcidID{0000-0002-4535-3956} and\\
Sabina D. Raducan\orcidID{0000-0002-7478-0148} and\\ 
Thomas Meier\orcidID{0000-0001-9682-8563} and\\ 
Martin Jutzi\orcidID{0000-0002-1800-2974} and\\ 
Joachim Stadel\orcidID{0000-0001-7565-8622} and\\
Ravit Helled\orcidID{0000-0001-5555-2652}
}
\institute{Christian Reinhardt \at Department of Astrophysics, University of Z{\"u}rich, Winterthurerstrasse 190, 8059 Z{\"u}rich, Switzerland, \email{christian.reinhardt@uzh.ch} \at Physics Institute, Space Research and Planetary Sciences, University of Bern, Sidlerstrasse 5, 3012 Bern, Switzerland, \email{christian.reinhardt@unibe.ch}
\and Sabina D. Raducan \at Physics Institute, Space Research and Planetary Sciences, University of Bern, Sidlerstrasse 5, 3012 Bern, Switzerland, \email{sabina.raducan@unibe.ch}
\and Thomas Meier \at Department of Astrophysics, University of Z{\"u}rich, Winterthurerstrasse 190, 8059 Z{\"u}rich, Switzerland, \email{thomas.meier5@uzh.ch}
\and Martin Jutzi \at Physics Institute, Space Research and Planetary Sciences, University of Bern, Sidlerstrasse 5, 3012 Bern, Switzerland, \email{martin.jutzi@unibe.ch}
\and Joachim Stadel \at Department of Astrophysics, University of Z{\"u}rich, Winterthurerstrasse 190, 8059 Z{\"u}rich, Switzerland, \email{joachimgerhard.stadel@uzh.ch}
\and Ravit Helled \at Department of Astrophysics, University of Z{\"u}rich, Winterthurerstrasse 190, 8059 Z{\"u}rich, Switzerland, \email{ravit.helled@uzh.ch}
}

\maketitle

\abstract{Impacts play a fundamental role in shaping the physical and chemical properties of the objects in our Solar System. Given the challenges in replicating such collisions through laboratory experiments, computer simulations are an important tool to investigate their outcomes. Accurately modelling material properties such as shear strength, porosity, and the formation of cracks is crucial for understanding impacts on small bodies like asteroids and comets. Very large and massive objects are dominated by self-gravity and can be approximated as a fluid. In this regime the equation of state used to model the behaviour of the constituent materials plays a key role. However, for bodies of several hundred kilometres, which are already spheroidal due to self-gravity, shear strength must still be considered. This impact regime is most challenging to model and therefore often overlooked in publications. In this review we present different impact regimes and the relevant physics that must be included. We then discuss their application to a variety of Solar System objects and assess how recent observations and numerical simulations, focussing on the Smoothed Particle Hydrodynamics method, can be used to inform our understanding of impact processes and solar system formation. }


\section{Introduction}\label{sec:introduction}
Collisions play a key role in the formation and evolution of planetary systems. In the early stages of planetary accretion, planetesimals, the smallest building blocks, form via streaming instability and grow by accumulating other planetesimals and pebbles. These collisions occur initially at low velocities and primarily result in merging of bodies. As some planetesimals grow more massive, they begin to gravitationally stir the surrounding smaller bodies, leading to increasingly energetic impacts. This transition reduces the efficiency of the accretion. The final stage of planet formation is dominated by giant impacts (GI), high-energy collisions between forming planets and remaining planetary embryos. The last few giant impacts a planet experiences sets the initial conditions for its interior evolution and cooling. This also has a large influence on orbital dynamics, physical and chemical composition, including the presence of volatiles and water, surface and atmospheric properties, as well as the formation of satellites. 

In the Solar System, several planetary features are attributed to GI, such as Mercury's large core \cite{benzOriginMercury2007, asphaug_mercury_2014, chau_forming_2018}, the formation of Earth's Moon \cite{Canup2001a, cuk_making_2012, canup_forming_2012, dengPrimordialEarthMantle2019, Timpe2023, meier_systematic_2024}, the Martian dichotomy \cite{marinova_mega-impact_2008, Emsenhuber2018, Ballantyne2023, cheng_combined_2024} and satellite system \cite{rosenblatt_accretion_2016, hyodo_impact_2017}, the lunar dichotomy \cite{jutzi_forming_2011, zhu_are_2019}, Jupiter's dilute core \cite{liuFormationJupitersDiluted2019, meierOriginJupitersFuzzy2025}, the Uranus-Neptune dichotomy \cite{kegerreis_consequences_2018, kurosaki_exchange_2018, reinhardtBifurcationHistoryUranus2020, woo_did_2022}, as well as the large gravitational anomalies associated with impact basins on Mercury and Pluto \cite{Ballantyne2024}. Once the GI phase is over, numerous small bodies composed of fragments of larger bodies or unaccreted planetesimals, remain. This marks the beginning of a prolonged period dominated by very energetic smaller-scale collisions, which continues to shape planetary surfaces to this day. Therefore, understanding the characteristics of the small body populations is vital not only for reconstructing the planets' past but also for assessing potential future threats for the Earth. 

The small bodies observed today (i.e., asteroids, comets, Kuiper Belt objects (KBOs)) are the product of billions of years of collisional evolution. While most large asteroids ($>\SI{100}{\kilo\meter}$) have survived largely intact since the early Solar System \cite{bottke_fossilized_2005, morbidelli_asteroids_2009}, smaller asteroids ($<\SI{50}{\kilo\meter}$) have much shorter collisional lifetimes \cite{bottke_fossilized_2005} and often originate from the breakup of larger parent bodies. Many of these fragments reaccumulate into gravitationally bound rubble-piles, as seen in asteroid families \cite{farinellaAsteroidFamiliesOld1996} and space mission observations \cite{fujiwara_rubble-pile_2006}. Their current shapes, structures, and compositions provide crucial insights into the history of planetary accretion and disruption processes. Beyond the asteroid belt, planetary migrations, particularly Neptune’s, have shaped the orbits and collisional history of KBOs.

These distinct impact phases in planetary system formation and evolution correspond to different impact regimes, determined by the size of the colliding bodies and the impact energy. In this chapter we explore the core physics governing impacts and the various regimes given by the size-scale and relative velocities, ranging from high-energy, high-velocity giant impacts, to low-velocity mergers between kilometer-sized objects. We highlight selected impact studies across different regimes that have advanced our understanding of Solar System evolution and collisional physics focussing on the Smoothed Particle Hydrodynamics (SPH) method \cite{gingold_smoothed_1977, lucy_numerical_1977}.

\section{Impact Physics}\label{sec:impact_physics}
As outlined in the previous section, planetary bodies in the Solar System have experienced countless low- and high-velocity impacts throughout their history. These collisions involve a large range of pressures, temperatures, and deformation regimes that may fundamentally alter the physical state of impacted material. Understanding the physics of impacts is therefore essential for reconstructing planet formation and evolution, interpreting planetary surface modifications, and understanding the collisional histories of small bodies.

\subsection{Impact Regimes}\label{sec:impact_regimes}

\begin{figure}[ht]
\centering
\includegraphics[width=\textwidth]{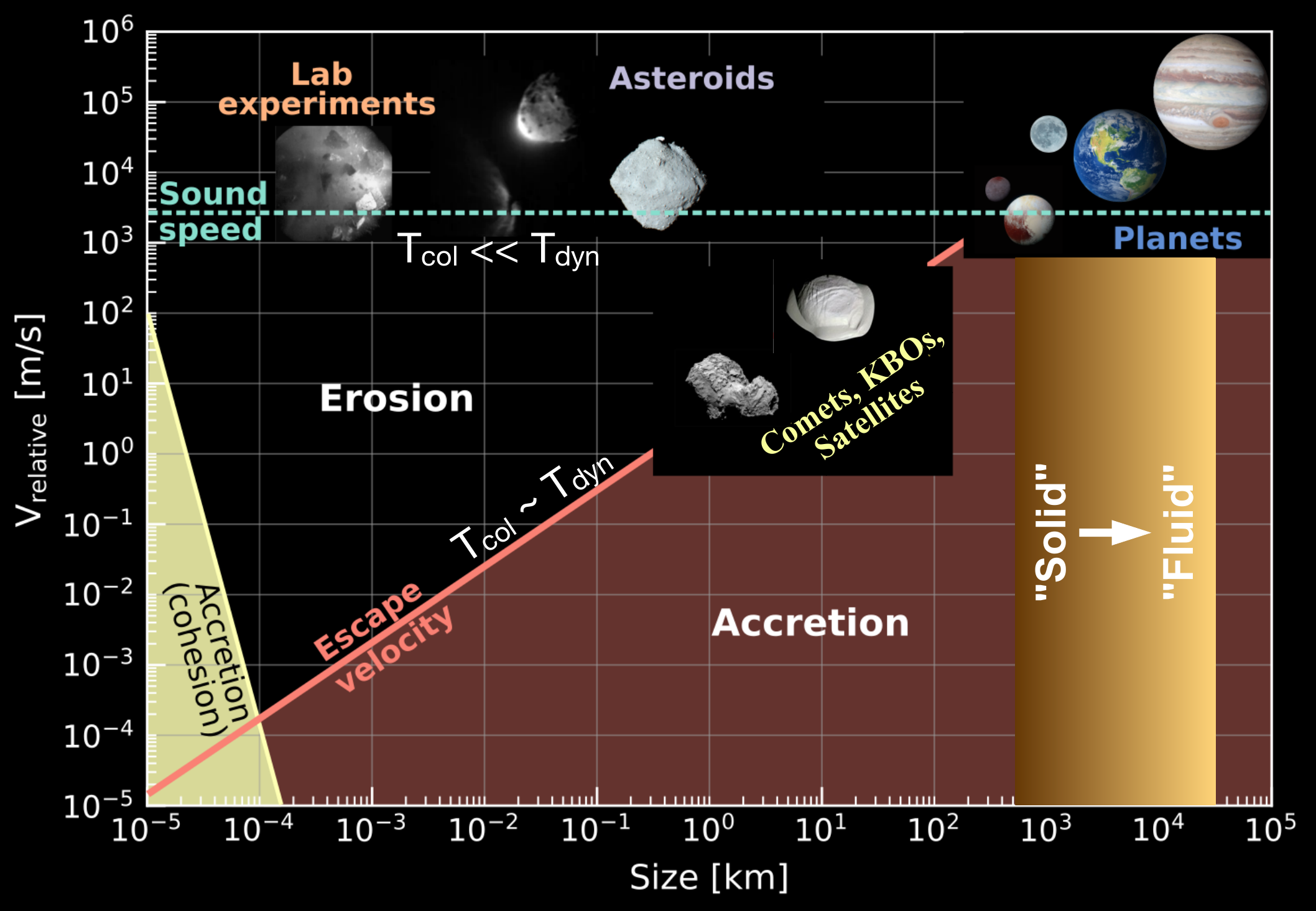} 
\caption{
Illustration of the different impact regimes discussed in Sect.~\ref{sec:impact_regimes}. Depending on the size of the target and the relative velocity, impacts can be divided into different regimes where different physical processes are relevant. For very small bodies, the collision time is typically much smaller than the dynamical time $\tau_{dyn} \sim 1 / \sqrt{G \rho}$ and gravitationally driven processes (material flow; ejecta reaccumulation) are decoupled from the initial impact phase (shock wave propagation). Depending on the impact velocity relative to both the mutual escape velocity of the colliding bodies and the sound speed of the involved materials, different processes dominate, constrained by either self-gravity or material strength, or both (see Sect.~\ref{sec:impact_regimes} for details). Examples of impact scenarios presented in Sect.~\ref{sec:modelling_studies} are indicated.}
\label{fig:coll_regimes}
\end{figure}

The conditions for impact processes in the Solar system span more than ten orders of magnitude in size and relative velocity of the involved bodies, making a general description very challenging. Depending on the size (or mass) of the colliding bodies and the impact velocity, collisions can be classified into different regimes (see Fig.~\ref{fig:coll_regimes} for an overview). Each regime is governed by distinct physical processes that shape the outcome of the impact \cite{asphaug_global_2015}.

There are several physical quantities that can be used to distinguish between the different impact regimes. These regimes can to some extent overlap which further complicates a description. One is the \textit{sound speed} (horizontal line in Fig.~\ref{fig:coll_regimes}) which is the velocity at which a sound wave propagates through a body. Typical impact velocities in planetary system are of the order of several \SI{}{\kilo\meter\per\second} and therefore often exceed the sound speed of the constituent materials which range from a few \SI{100}{\meter\per\second} in granular material to several \SI{}{\kilo\meter\per\second} in rock). Such \textit{hypervelocity impacts} initiate shocks in the colliding bodies and result in extreme pressures and temperatures that produce irreversible changes of the physical state of the involved materials \cite{meloshImpactCrateringGeologic1989}. Interaction of the shock wave with free surfaces, e.g., at the opposite surface of the impactor, result in a rarefaction wave that is reflected back into the shocked material decompressing it to ambient conditions conserving entropy. Shock release can result in melting or vaporization and accelerate material to very high velocities.

Another important quantity is the mutual \textit{escape velocity} of the colliding bodies which marks the boundary between \textit{accretion} and \textit{erosion} (diagonal line in Fig.~\ref{fig:coll_regimes}). Typical values for the escape velocity can vary by several orders of magnitude depending on the mass and composition of the bodies. For example, while the escape velocity of a \SI{10}{\kilo\meter} sized planetesimal is a few \SI{}{\meter\per\second}, for Earth it is \SI{11}{\kilo\meter\per\second}. While the detailed outcome of collisions also depends on the impact geometry, for impact velocities on the order of or below the mutual escape velocity, they typically result in merging and net \textit{accretion}. On the other hand, for velocities greater than a few times the escape velocity, impacts typically lead to net \textit{erosion}. If the projectile-to-target size (and mass) ratio is very small an impact results in cratering. Such impacts usually lead to little mass loss and the effect is limited to a well defined geometrical locus \cite{asphaug_global_2015}. For larger projectiles that are comparable in size (and mass) to the target, impacts affect the target on a global scale and transition from cratering to erosion or disruption of the involved bodies (e.g., \cite{asphaug_global_2015}). Giant impacts during the last stage of planet formation involve very massive bodies and high velocities, and examples of such global scale impacts are presented in Sect.~\ref{sec:giant_impacts} and~\ref{sec:global_impacts}.

Another important aspect are the relevant \textit{time scales} of the problem. If the impact velocity is close to the mutual escape velocity (diagonal line in Fig.~\ref{fig:coll_regimes}) then the collision time $\tau_{coll}$ is comparable to the dynamical time $\tau_{dyn} \sim 1 / \sqrt{G \rho}$. However, for the impact conditions that are accessible by laboratory experiments and are typical in the asteroid belt, i.e., for small bodies and high impact velocities (top left corner of Fig.~\ref{fig:coll_regimes}), collisions occur within a few seconds and shock propagation is decoupled from gravitational reaccumulation. While this simplifies the modelling of the initial impact phase, where self-gravity can be neglected, it requires advanced hybrid-schemes to model the entire collision process that includes gravitationally driven material flow and ejecta reaccumulation (see Sect.~\ref{sec:small_impacts} for examples of such collisions).

For small bodies, \textit{material strength} dominates over gravity, which is reflected in their highly irregular shapes. Impact processes in this regime are constrained by material rheology. When impact-induced stresses are comparable to the material strength (shear, tensile, or compressive), mechanical failure and irreversible processes -- such as fracturing, crushing, and frictional dissipation -- can occur. This regime also includes cratering impacts on planetary surfaces, i.e., all situations where the relevant stresses, including gravitational overburden pressures, are comparable or below the maximum yield strength of the involved material.

For larger, planetary-scale bodies, the interior pressure due to gravity exceeds the maximum yield strength, making self-gravity the dominant force and causing the bodies to become spherical. This marks the transition (and sets the size threshold) at which the impact response of an object -- on a planetary scale -- can be approximately described by a \textit{fluid rheology}. In this regime the equation of state is most critical (see Sect.~\ref{sec:equation_of_state} for details) for the accurate modelling of shock compression and release, while the effects of material strength are more subtle, but may still influence the speed and the decay rate of shock waves. At intermediate scales (e.g., Mars-, Mercury-, and Moon-sized bodies), both material strength and self-gravity are equally important and have to be accounted for at the same time (e.g., \cite{Emsenhuber2018, Ballantyne2023, Ballantyne2024, denton_capture_2025}), making these collisions particularly challenging to model.

\subsection{Numerical Modelling}\label{sec:numerical_modeling}
Laboratory experiments provide valuable insights into impact physics, but are inherently limited in scale, impact velocity, and the ability to incorporate self-gravity. Therefore, numerical simulations are required in order to study impact processes across a wide range of conditions, spanning the diverse conditions encountered, from giant impacts to laboratory-scale cratering. Computer codes designed to model such impacts are often referred to as hydro-codes since early codes treated materials as fluids, mostly focusing on bodies subjected to strong shocks. However, modern shock physics codes incorporate sophisticated material models, including strength, friction, porosity and failure models, making them appropriate for studying both large and small-scale impacts \cite{pierazzoValidationNumericalCodes2008, jutziModelingAsteroidCollisions2015}. 

Numerical modelling of impact processes is based on three fundamental pillars. The first are Newton's laws of motion which describe the fluid motion due to pressure gradients and, if relevant, gravity. This is usually achieved by solving the equations of continuum mechanics via the conservation of mass, momentum, and energy. These equations can be implemented either in a stationary coordinate system (Eulerian methods) or in one that is moving with the material (Lagrangian methods). The conservation equations are material independent and the specific properties of a material such as the response to stress are described by the other two pillars. The second pillar is the equation of state which relates the pressure to the density and internal energy, and also corresponds to the volumetric component of stress. It is most relevant for very high energy impacts and during the early stages of cratering when the pressures exceed material strength. The last pillar is constitutive (or rheological) models that describe the materials deformation (strain) due to the deviatoric component of stress. While the physics governing continuum mechanics and gravity are well understood, research regarding improvements of the numerical methods to solve these equations, the equation of state and rheological models of planetary materials is still a very active area of research. 

\subsubsection{Smoothed Particle Hydrodynamics and Advances in Numerical Modelling}
Over the last decades several numerical methods to solve the conservation equations have been developed (e.g., \cite{pierazzo_brief_2004, pierazzoValidationNumericalCodes2008, liu_smoothed_2010}). These can be broadly divided into grid-based and mesh-less techniques. The most prominent meshless technique in impact modelling is the Smoothed Particle Hydrodynamics (SPH) method \cite{monaghan_smoothed_1992, springel_smoothed_2010, priceSmoothedParticleHydrodynamics2012} which samples the fluid with particles that evolve with the flow. Fluid quantities, such as the local density, temperature and pressure and their derivatives, are then computed from neighbouring particles using smoothing kernels. Being a particle-based Lagrangian method SPH has no advection errors, can model large deformations and free surfaces often encountered in impact simulations and tracking the history and origin of the material is trivial. However, it is difficult to implement arbitrary boundary conditions and depending on the SPH formulation the method can exhibit substantial errors at contact discontinuities such as material interfaces and free surfaces. Furthermore, the method tends to suppress (physical) mixing between different materials \cite{agertz_fundamental_2007, dengEnhancedMixingGiant2019} and cannot model variable mixtures of different materials \cite{meierOriginJupitersFuzzy2025}.

Since the pioneering work of W.Benz et al. adapting SPH for impact modelling \cite{benz_origin_1986, benz_origin_1987, benz_origin_1989} the method has substantially improved both in terms of computational efficiency, numerical accuracy and physical realism. Early simulations were limited to a few thousand particles which, for example, hampered efforts to identify a successful Moon-forming impact because the disk was resolved with only a few particles and one iron particle contained more mass than the upper estimated limit of the iron mass fraction \cite{canup_dynamics_2004}. Since then increases in computing power and efficient parallelization has lead to an increase in particle numbers to $10^9$ particles in recent simulations \cite{meierOriginJupitersFuzzy2025}. While some properties, such as the bound mass, can be faithfully predicted with only a few $10^5$ particles, the mass and angular momentum distribution can vary even if more than $10^7$ particles are used \cite{hosono_unconvergence_2017, kegerreisImmediateOriginMoon2022, meier_systematic_2024}. Modelling low mass structures like the Earth's crust, oceans, or atmosphere in global scale collisions requires very high particle numbers and is extremely challenging even with state-of-the-art simulations. In smaller-scale cratering impacts, resolving the ejecta size distribution or the detailed surface morphology, including boulders and other small-scale heterogeneities, is similarly challenging (Section \ref{sec:outlook}). 

Since its inception SPH has undergone several advancements in numerical accuracy and physical realism. Most important for impact modelling are the more realistic treatments of free surfaces \cite{reinhardtNumericalAspectsGiant2017} and material interfaces \cite{hosono_giant_2016, reinhardtBifurcationHistoryUranus2020, ruiz-bonillaDealingDensityDiscontinuities2022} which substantially reduces artificial surface tension inhibiting mixing (also see Section \ref{sec:giant_impacts_moon}). In parallel, progressively more realistic material models incorporating tensile fragmentation \cite{Benz1994,Benz1995}, porosity \cite{Jutzi2008}, friction \cite{Jutzi2015}, and the presence of boulders \cite{raducanDART2024} have been developed and applied to the study of small body collisions (see also Sections \ref{sec:material_rheology} and \ref{sec:small_impacts}).

\subsubsection{The Equation of State}\label{sec:equation_of_state}
An equation of state (EOS) describes the thermodynamic state of a material by relating pressure, density and temperature (as well as other thermodynamical parameters such as entropy and internal energy). For modelling impacts the EOS of planetary materials, such as rocks, ices or metals, must capture a wide range of pressures and temperatures, from ambient conditions to those experienced deep within a planetary interior or during a giant impact (up to several hundreds of GPa). The EOS is unique for each material and provides a relationship between the thermodynamic state variables. Together with the conservation equations it determines how much energy is converted into heating, melting, and vaporization, as well as how shock pressures decay as waves propagate outward from the point of impact. Specifically for modelling hypervelocity impacts the EOS must accurately capture the material's Hugoniot curve which corresponds to the peak pressures, densities and temperatures at the shock front and determines the initial state for shock release \cite{stewart_shock_2020}. However, developing accurate EOS models is challenging due to the difficulty of obtaining experimental data at the immense pressures and temperatures of giant impacts. Laboratory experiments, while useful, can only replicate a narrow range of these conditions. Additionally, the transient nature of such phenomena complicates direct measurement of key properties like density and temperature. As a result, EOS models often rely on theoretical approximations and extrapolation to fill these gaps.

Two widely used EOS for modelling impact processes are the Tillotson EOS and ANEOS. The Tillotson EOS \cite{tillotsonMetallicEquationsState1962} (see \cite{meloshImpactCrateringGeologic1989, brundage_implementation_2013} for a more detailed description) provides a simple analytic description of typical planetary materials that accurately models shocks and covers a wide range of densities and internal energies. However, the good agreement with data from shock experiments is limited to relatively low pressures and the Tillotson EOS lacks a realistic model for the expanded states, e.g., the liquid/vapour phase, which results in substantial inaccuracies during shock release making it a questionable choice to model strong shocks \cite{wissing_new_2020, stewart_shock_2020, meierEOSResolutionConspiracy2021}. Furthermore, the EOS it not thermodynamically complete and the temperature has to be estimated assuming a constant heat capacity, usually determined at the reference state, which is a rather crude approximation (e.g., \cite{brundage_implementation_2013}).

ANEOS (ANalytic Equation Of State) \cite{thompson_m-aneos_2019, melosh_hydrocode_2007, stewart_shock_2020} addresses several short-comings of the Tillotson EOS, especially for modelling very energetic collisions that result in strong shocks and substantial melting or vaporization of the material. It is based on fitting analytic expressions of the Helmholtz free energy in different regions of the EOS to experimental data and provides additional thermodynamic data such as temperature and entropy. ANEOS is a thermodynamically complete and consistent model of the different phases including phase transitions and mixed phases. More recent improvements account for multi-atomic vapour \cite{melosh_hydrocode_2007} and a more accurate treatment of the thermal part of the free energy in the liquid phase \cite{stewart_shock_2020} resulting in an excellent fit to experimental Hugoniot data up to \SI{1000}{\giga\pascal}. Nonetheless, ANEOS also has several limitations. The set of materials implemented in the EOS are limited. In its current formulation only two phase transitions, e.g., melt and a high pressure solid-solid transition, can be simultaneously modelled \cite{collins_improvements_2014}. Because relatively simple expressions are used to describe each phase, the accuracy of the model in certain regions of the phase diagram, e.g., the vapour phase, can be poor. Finally, ANEOS and other more sophisticated analytic EOS are computationally expensive and therefore precomputed tables are used in simulations which can introduce large interpolation errors, especially at phase transitions, and violate thermodynamic consistency.

\subsubsection{Material Rheology and Strength}\label{sec:material_rheology}
\begin{figure}[ht]
\centering
\includegraphics[width=\textwidth]{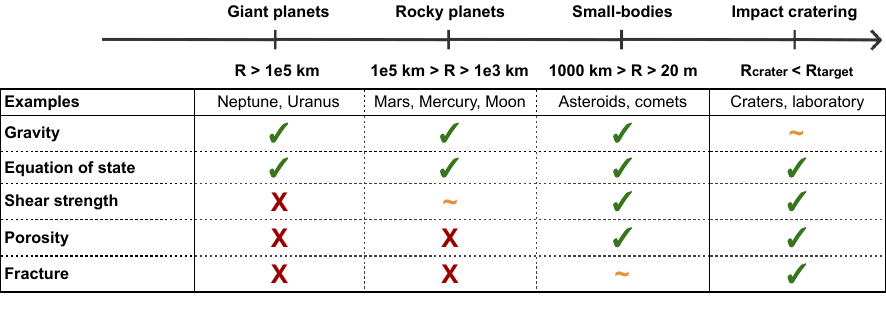} 
\caption{Figure highlighting the relevant physics for global-scale impacts and impact cratering. The symbols indicate which physics are relevant in a specific regime (tick: important, tilde: can be relevant, cross: not important).}
\label{fig:impact_scales}
\end{figure}

As described in Sect.~\ref{sec:impact_regimes}, material strength plays a fundamental role in various impact regimes (also see Fig.~\ref{fig:impact_scales}.) Material rheology describes how a material responds to stresses that induce deformation, including elastic, plastic, and viscous behaviour. The critical stress at which a material undergoes permanent deformation is defined as its strength, which varies depending on the type of stress: compression, tension, or shear. In most geologic materials, compressive strength is the highest, tensile strength is the lowest, and shear strength (or yield strength) is pressure-dependent. The simplest constitutive models for material deformation include elastic (where stress is linearly proportional to strain, and deformation is fully reversible), Newtonian fluids (where stress is proportional to the strain rate, describing viscous flow) and plastic materials (which deform elastically up to a yield stress, beyond which they flow without additional resistance) \cite{jutziModelingAsteroidCollisions2015}. Rocks and geological materials are often modelled as pressure-dependent plastic materials \cite{lundborg_strength-size_1967, Collins2004,Jutzi2015}, where yield strength increases with pressure, accounting for friction. This behaviour reflects how real planetary materials respond under different loading conditions.

In addition to strength, porosity is a key factor influencing impact outcomes, particularly in low-density materials such as asteroids and comets. Highly porous materials absorb impact energy through pore collapse and compaction, significantly altering shock wave propagation. Commonly used models for simulating porous materials in shock physics codes are the $P-\alpha$ \cite{Jutzi2008, Jutzi2009} and $\epsilon-\alpha$ models \cite{Wunnemann2006}, which describe how porosity evolves under compression. In porous bodies, impacts result in energy dissipation and reduced ejecta velocities, leading to different crater morphologies and impact structures compared to solid bodies. While impact compaction leads to a reduction of porosity, impact-induced fracturing and shear failure tend to increase it \cite{Collins2014_dilatancy}.

Porosity also affects collisional fragmentation thresholds, with highly porous materials requiring higher impact energies to be disrupted compared to non-porous materials. Porosity compaction has significant implications for the impact history of rubble-pile asteroids, comets, and other icy bodies in the Kuiper Belt, where porosity plays a major role in determining collisional evolution.

During high-energy impacts, target materials experience fracturing, leading to extensive damage and a reduction in strength \cite{Benz1994, Benz1995}. Once fractured, rock behaves as a granular material, where strength is primarily dictated by frictional resistance rather than intrinsic strength \cite{Jutzi2015}. In the fragmentation regime, impact outcomes depend on factors such as pre-existing fractures, material heterogeneity, and dynamic loading conditions. At small scales, fragmentation plays a dominant role in impact processes, especially in the case of rocky asteroids or boulders of rubble-pile asteroids. In such bodies, large portions of the impact energy can be absorbed by fracturing individual boulders, as opposed to crater excavation or target disruption \cite{Raducan2024_lessons}.

\section{How Impact Modelling has Advanced the Study of the Solar System}\label{sec:modelling_studies}
The impact modelling community has extensively used different shock physics codes to investigate impact processes across the Solar System, spanning a broad range of conditions and regimes, from giant planetary-scale collisions to crater formation and small-body impacts on asteroids and comets. In this section, we highlight key contributions in the different impact regimes that have advanced our understanding of the role of impact processes in the Solar System.

\subsection{Giant Impacts}\label{sec:giant_impacts}
Solar System formation models predict that large impacts were common during the planetary accretion \cite{agnor_character_1999, chambers_making_2001, quintana_frequency_2016} (and also see Chapt.~20). Since the involved planets are very massive and the impact velocities very high, many of these collision are in a regime where the fluid approximation is valid (upper right corner of Fig.~\ref{fig:coll_regimes}), material strength can be neglected and the evolution is governed by shock physics, the EOS and self-gravity. These high-energy collisions play a key role in shaping the final planets and can provide critical constrains on planet formation in the Solar system (e.g., \cite{fang_moon-forming_2025}).

\subsubsection{The Uranus-Neptune Dichotomy}
\begin{figure}[t!]
	\centering
	\includegraphics[width=0.9\linewidth]{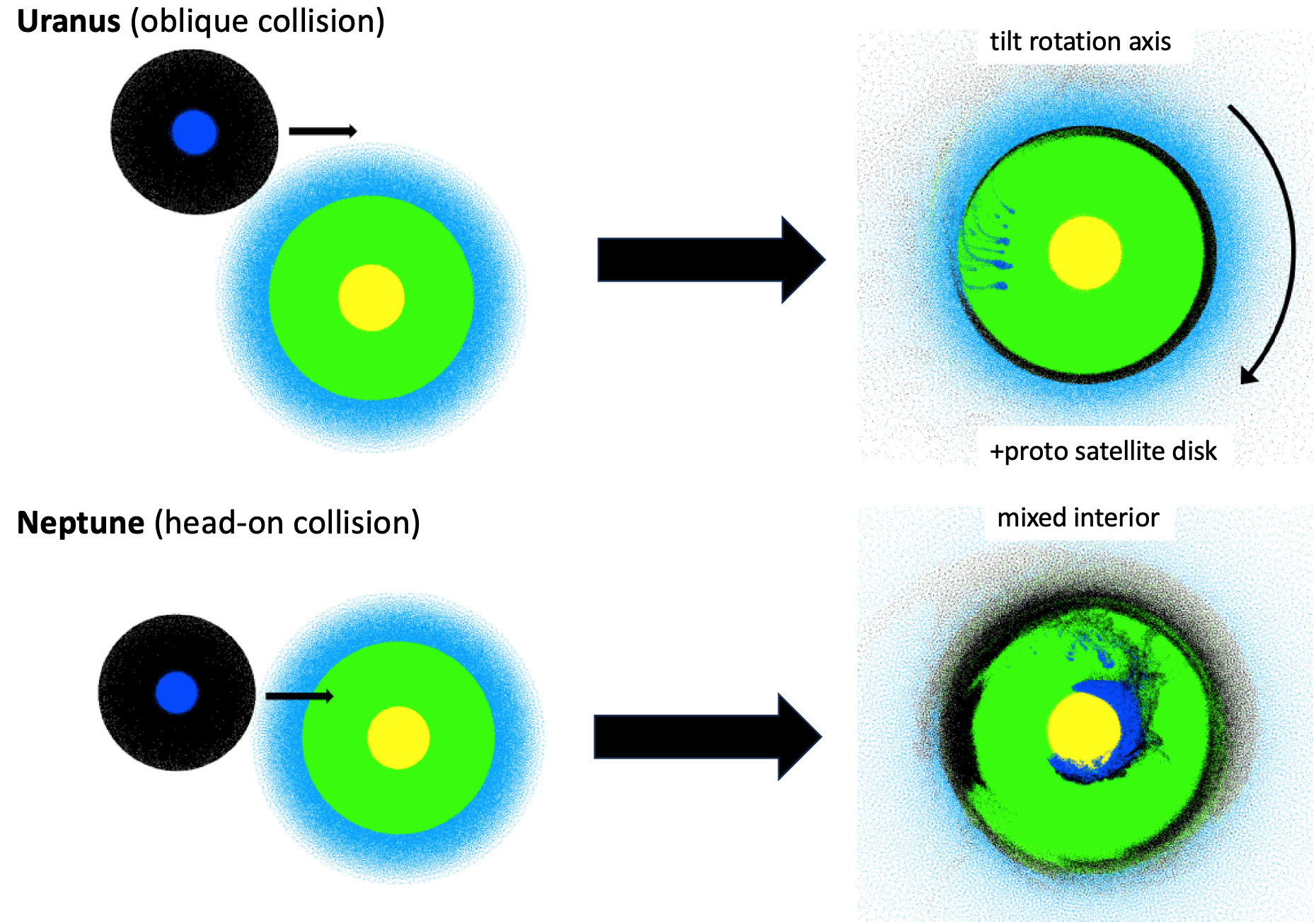}
	\caption{Schematic representation of the role of giant impacts in the dichotomy between Uranus and Neptune (not to scale). An oblique giant impact on Uranus could have tilted its spin axis while ejecting material to form a disk and its regular satellites, preserving internal stratification. For Neptune, a nearly head-on collision may have mixed its interior, creating a near-adiabatic thermal profile that can explain its rapid cooling. Figure adapted from \cite{reinhardtBifurcationHistoryUranus2020, Helled2025}.}
 \label{fig:uranus_neptune}
\end{figure}

An open question in planetary science relates to the differences between the two ice giants, Uranus and Neptune. Despite the similarities between the planets, such as their similar masses, mean densities, and orbital distances, they also exhibit striking differences in their physical and orbital properties. Uranus is tilted by $\sim \SI{97}{\degree}$ with respect to the ecliptic plane, while Neptune's tilt is only $\sim \SI{28}{\degree}$ (more typical for Solar System planets). Their satellite systems also differ: Uranus’ major moons follow regular, prograde orbits, consistent with accretion from a circumplanetary disk, whereas Neptune’s largest moon, Triton, follows an irregular, inclined orbit, likely being a captured object. Finally, while Uranus seems to be in thermal equilibrium with Solar insolation, Neptune emits nearly ten times more heat, indicating differences in heat transport. Since formation models suggest that both planets experienced at least one giant impact \cite{izidoro_accretion_2015, valletta_possible_2022, eriksson_can_2023, esteves_accretion_2025}, these differences are often attributed to different impact histories \cite{stevensonUranusNeptuneDichotomyRole1986, podolak_what_2012} (as depicted in Fig.~\ref{fig:uranus_neptune}).

\begin{figure}[ht]
\centering
\includegraphics[width=\linewidth]{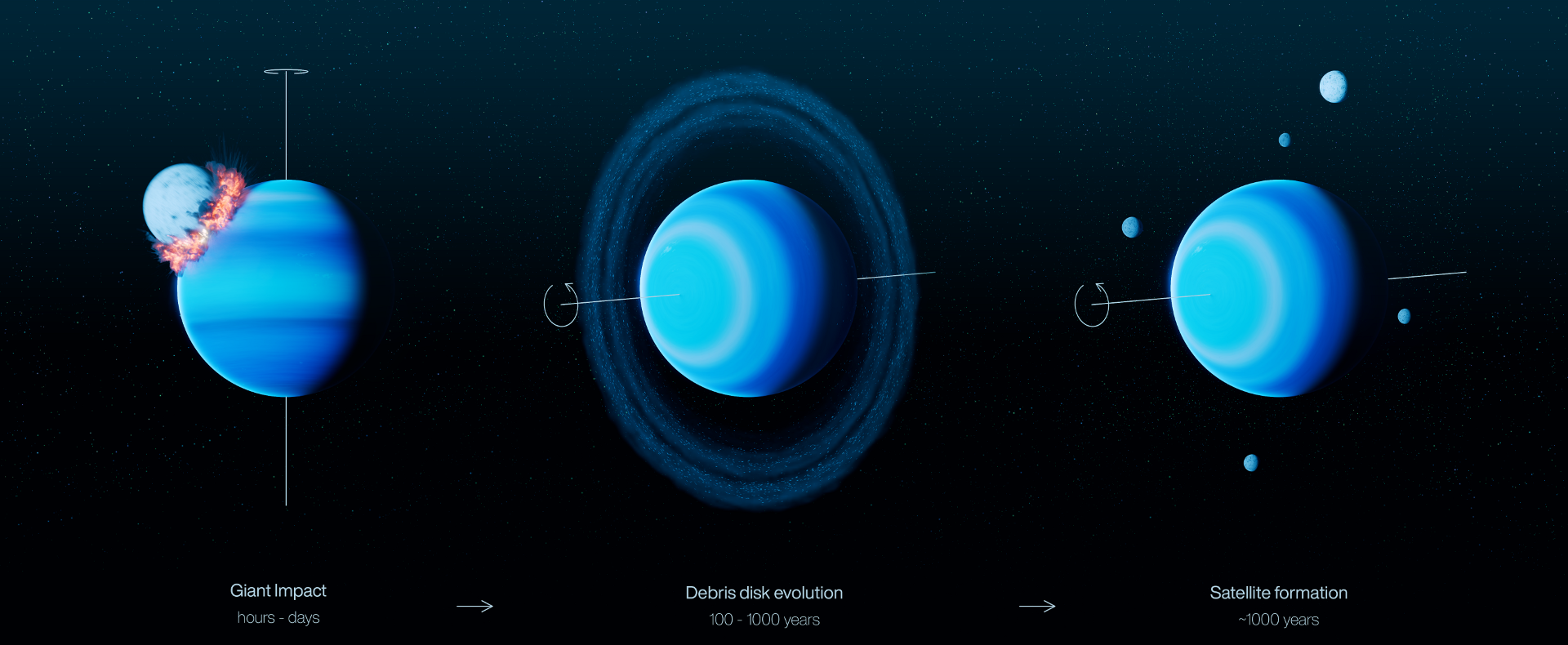}
\caption{Schematic representation of the formation of Uranus' five regular moons from a giant impact \cite{woo_did_2022}. A massive collision ejects material into Uranus' orbit forming a disk which viscously spreads and cools. Because of their very different condensation temperatures the rock-to-ice ratio of the disk changes from ice rich to $\sim 1$. The condensed material then forms moonlets that merge and form the Uranian moons.}
\label{fig:uranus_moons}
\end{figure}

Impact simulations \cite{reinhardtBifurcationHistoryUranus2020} show that a grazing collision can explain Uranus' large axial tilt and form a circum-planetary disk, while having little effect on the planets deep interior. For significant material mixing, the impactor's core must be tidally stripped into smaller fragments in a graze-and-merge collision. A later study linked Smoothed Particle Hydrodynamics (SPH) impact simulations with accretion models, showing that the resulting disk could produce satellite systems matching Uranus' moons in mass, size, and orbital properties \cite{woo_did_2022} (see Fig. \ref{fig:uranus_moons}). In contrast, a head-on collision for Neptune leads to mass accretion and interior heating, while ejecting little mass into the orbit, which is consistent with Triton's irregular orbit.

The extent to which these collisions align with planet formation models remains unclear. Recent studies propose that Uranus and Neptune formed from massive planetary embryos that accumulated beyond Saturn's orbit, which acted as a dynamical barrier against inwards migration \cite{izidoro_accretion_2015}. In this scenario, high-mass embryos underwent one to three giant impacts, forming Uranus and Neptune analogues. However, 3D SPH simulations \cite{chau_could_2021} suggest that reproducing their obliquities and angular momentum remains challenging. Additionally, large impactors in this scenario tend to create overly massive proto-satellite disks, inconsistent with observations. In order to fully understand the role of giant impacts in the early evolution of Uranus and Neptune, further integration of impact simulations with formation and evolution models is required.

\subsubsection{The Formation of the Earth-Moon System}\label{sec:giant_impacts_moon}
The most prominent example of a giant impact (GI) in the Solar System is the formation of the Earth-Moon system. The leading hypothesis proposes that the early Earth experienced a GI resulting in a circum-planetary debris disk from which the Moon accreted \cite{Hartmann1975, Cameron1976}. The canonical version of this hypothesis states that the impactor was Mars-sized and that the impact occurred at an oblique angle and at a velocity close to the mutual escape velocity. The canonical model stands out, as it can simultaneously reproduce the total angular momentum of the Earth-Moon system and the low iron content of the Moon with impact conditions that are not especially unlikely. Early simulations \cite{benz_origin_1989, Canup2001a, Canup2004} suggest that in such an impact, approximately one lunar mass of iron-depleted material is placed in Earth's orbit.

However, this scenario has been challenged by geochemical data that show a striking similarity between lunar rocks and the Earth's mantle (e.g., \cite{wiechert_oxygen_2001, touboul_late_2007, zhang_proto-earth_2012}). In the canonical model, most of the material forming the proto-lunar disk is derived from the impactor. To reconcile this with the isotopic similarity, this would require that either both bodies had a very similar isotopic composition or strong mixing between the Earth and the proto-lunar disk occurred post-impact \cite{pahlevanEquilibrationAftermathLunarforming2007}. While the former scenario is possible, it requires a low probability event \cite{kaib_feeding_2015, mastrobuono-battisti_composition_2017}. The latter requires orders of magnitude longer timescales to equilibrate the two reservoirs than is suggested for the formation time of the Moon \cite{kokuboEvolutionCircumterrestrialDisk2000}. To resolve these inconsistencies, alternative scenarios were proposed: a merger between two half-Earth mass bodies \cite{canup_forming_2012} and a high-velocity impact on a rapidly counter-rotating Earth \cite{cuk_making_2012}, both resulting in excess angular momentum of one to two times the current value of the Earth-Moon system which requires a post-impact process to reduce it to the observed value. \cite{kegerreisImmediateOriginMoon2022} identified a scenario, where the Moon is formed directly as a post-impact satellite of an impact close to the canonical model. A recent systematic study of possible Moon-forming impact scenarios with pre-impact rotation by \cite{Timpe2023,meier_systematic_2024} found that the canonical scenario does not lead to a sufficiently heavy proto-lunar disk and that generating a post-impact Earth that has a similar impactor fraction as the disk requires near equal mass collisions. But this study identified a new scenario, where a rotating proto-Earth suffers an impact by a body that is roughly 3 times the mass of Mars, that is uniquely capable of producing a sufficiently heavy, iron-poor disk with an angular momentum budget close to the current value.

Most work in this field has been done using SPH to model the collisions. However, SPH suppresses mixing due to pressure errors at material interfaces and excessive damping of sub-sonic turbulence \cite{agertz_fundamental_2007,dengEnhancedMixingGiant2019}. Deng et al. \cite{dengEnhancedMixingGiant2019} demonstrated that the Meshless Finite Mass method significantly enhances mixing compared to SPH. They showed that the resulting inefficient angular momentum transport can lead to compositional stratification of Earth’s mantle following the canonical Moon-forming impact: while the upper mantle mixes well with impactor material and resembles the proto-lunar disk, the lower mantle remains pristine target material. If preserved over billions of years, this stratification naturally explains both the isotopic similarity between Earth and the Moon and the mantle’s heterogeneity. Even though a lot of work has been done in this regard, it is clear that the formation of the Moon is very much an unsolved problem.

\subsubsection{Forming Mercury via Giant Impacts}\label{sec:mercury}
Mercury, the innermost and smallest of the terrestrial planets, has a high mean density which suggests that it has a large iron core containing $\sim \SI{70}{\percent}$ of its mass \cite{hauck_curious_2013}. This composition cannot be explained by in situ formation models which would lead to a more Earth-like composition \cite{wurm_photophoretic_2013, lewis_metalsilicate_1972, aguichine_rocklines_2020}. One possibility is that Mercury experienced a very violent collision that stripped part of its mantle after formation \cite{wetherill_occurrence_1985}. Several impact scenarios have been proposed: a single giant impact with another planetary embryo \cite{benzOriginMercury2007}, a hit-and-run encounter with a more massive planet \cite{asphaug_mercury_2014} and multiple, smaller impacts \cite{chau_forming_2018}. Chau et al. \cite{chau_forming_2018} investigated these different impact scenarios in one numerical framework using SPH simulations. It was found that for the single impact scenario very high impact velocities are required and the parameter space of successful collisions is narrow. They showed that while the assumed composition of the impactor affects the post-impact mass of Mercury (by up to \SI{25}{\percent}) it does not affect the inferred total iron mass fraction. If Mercury's mantle was stripped in a hit-and-run collision, lower impact velocities are required, and the parameter space for successful impacts is somewhat larger. However, several collisions were highly disruptive and Mercury was destroyed in the collision. Also, the fate of the impactor (i.e., the more massive planet) in such an encounter remains unknown and should be investigated in future studies. Finally, multiple impacts of smaller bodies can strip Mercury's mantle and require less fine tuning in the impact conditions, but are constrained by the timing as successive impacts must occur within a relatively short interval.

Overall, explaining Mercury's composition and structure via a GI is difficult and requires rather specific impact conditions. Another challenge for these scenarios is the removal of the ejected material from Mercury's orbit in order to avoid re-accretion after several orbits. A more recent study on forming iron-rich exoplanets via GI confirms that unique circumstances are needed, which is consistent with the low frequency of observations \cite{reinhardt_forming_2022}. Clearly, additional work on the accretion of Mercury and other iron-rich planets linking impact simulations to formation models is required to understand this special class of planets. 

\begin{figure}[ht]
\centering
\includegraphics[width=\textwidth]{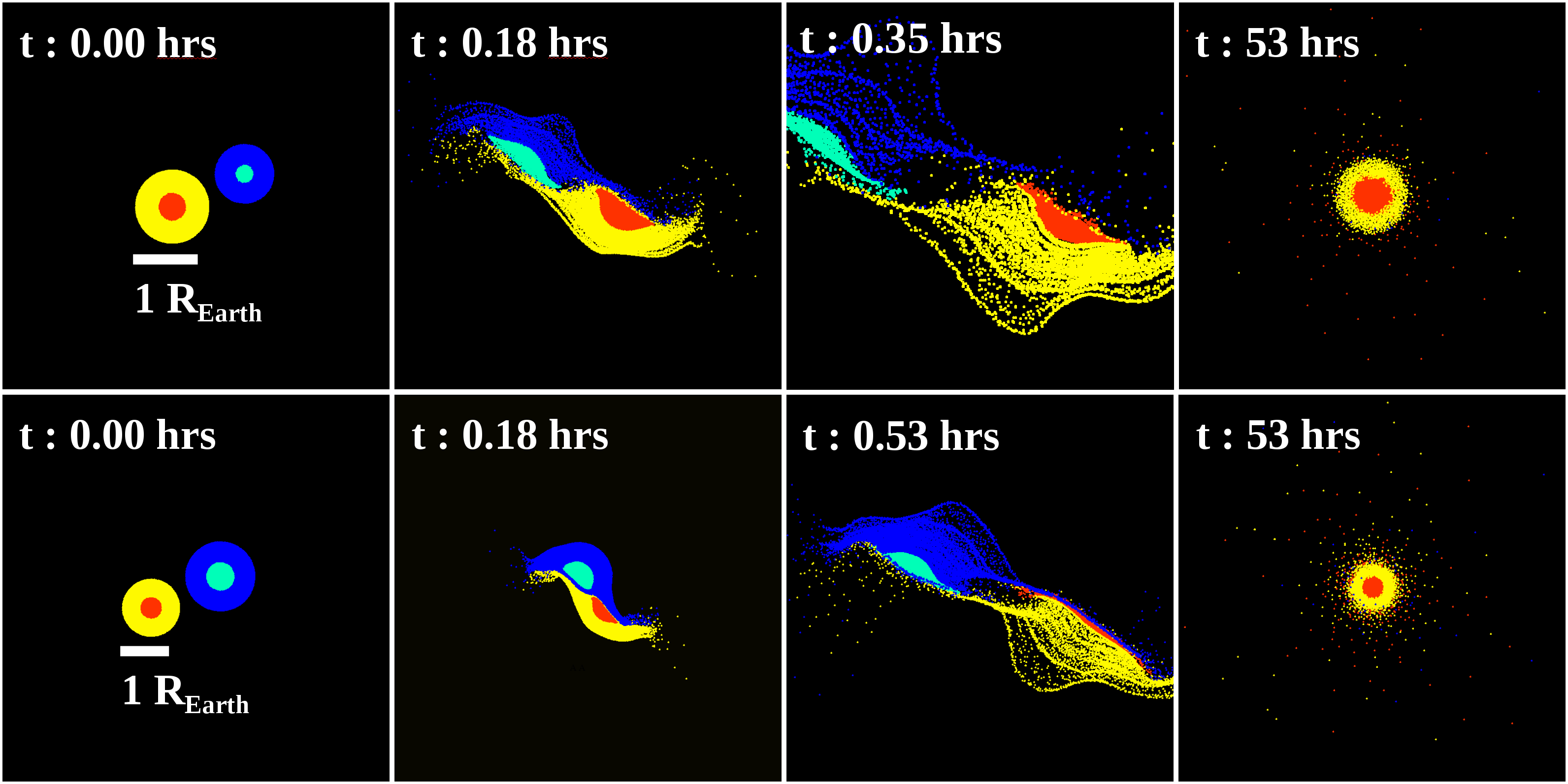} 
\caption{Two examples of giant impacts that can strip Mercury's mantle and explain it's large iron core. Shown are snapshots at different times where the figure size is kept constant. \textit{Top:} Proto-Mercury of \SI{2.25}{\MM} where \SI{}{\MM} is the mass of Mercury (red: core; yellow: mantle) collides with the impactor of \SI{1.125}{\MM} (turquoise: core; blue: mantle) at an impact parameter of $b = 0.5$ and $v = \SI{30}{\kilo\meter\per\second}$.
\textit{Bottom:} Proto-Mercury of \SI{2.475}{\MM} collides with an impactor of \SI{4.53}{\MM} at an impact parameter of $b = 0.5$ and $v = \SI{20}{\kilo\meter\per\second}$. Figure taken from \cite{chau_forming_2018}.}
\label{fig:mercury}
\end{figure}

\subsection{Global-scale Impacts}\label{sec:global_impacts}

Global-scale impacts are less energetic than the giant impacts discussed above but can still affect the target as a whole, leading to global-scale surface dichotomies, giant basins with sizes comparable to the planet's diameter and other distinct large surface features. This is a very challenging impact regime because self-gravity, accurate EOS and detailed material models are important (something often overlooked in prior studies).

\subsubsection{Mars' Surface Dichotomy}
\begin{figure}[ht]
\centering
\includegraphics[width=\textwidth]{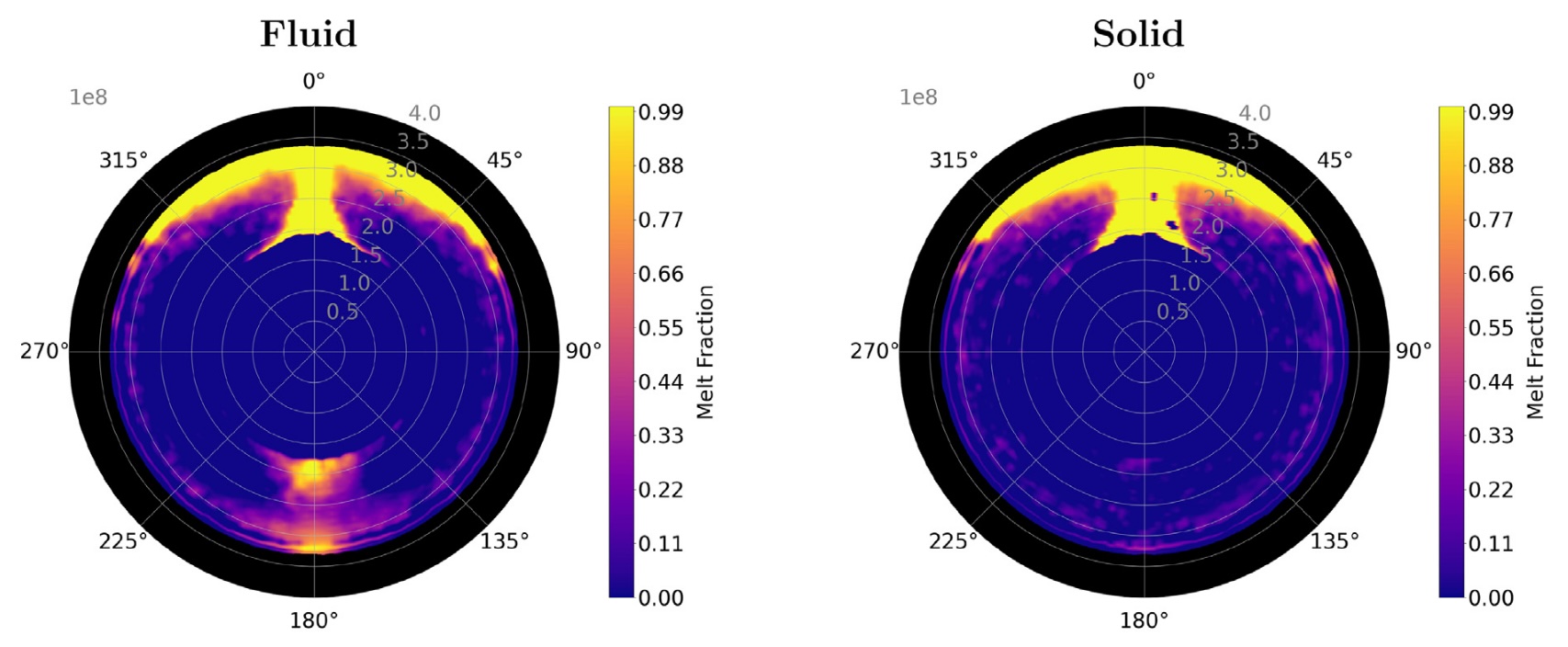} 
\caption{Melt-fraction distribution as a result of a global-scale impact on Mars. Left: fluid rheology; right: solid rheology. Figure taken from \cite{Ballantyne2023}.}
\label{fig:mars_melt}
\end{figure}

A global-scale impact is widely considered a key explanation for the stark differences in elevation and crustal thickness between Mars' two hemispheres, known as the "Martian Dichotomy". Initially, this hypothesis proposed that a massive impact in the northern hemisphere created the Borealis Basin \cite{marinova_mega-impact_2008}. However, more recent research suggests a hybrid model \cite{golabekOriginMartianDichotomy2011}, where the dichotomy results from impact-induced crust production rather than a single, large-scale basin. Ballantye et al. \cite{Ballantyne2023} employed a comprehensive set of SPH simulations that account for material strength, coupled with an advanced geophysical model of crust formation, to evaluate whether a giant impact in either hemisphere could explain Mars' observed crustal distribution. The modelling suggest that the traditional Borealis-forming impact scenario is unlikely due to excessive crust production and strong antipodal effects that are inconsistent with the present-day Martian southern hemisphere. Instead, the results support a scenario in which an impact in the southern hemisphere led to a localized magma ocean, which subsequently crystallized into a thicker crust than in the north. The study also confirmed the importance of including realistic material rheologies \cite{Emsenhuber2018} in this impact regime (Fig.~\ref{fig:mars_melt}). Subsequent long-term coupled SPH-thermochemical modelling allowed for refined assessment of the impact scenarios and comparison to today's surface features (\cite{cheng_combined_2024}; Chapt.~26).

\subsubsection{Pluto's Heart}
\begin{figure}[ht]
\centering
\includegraphics[width=\textwidth]{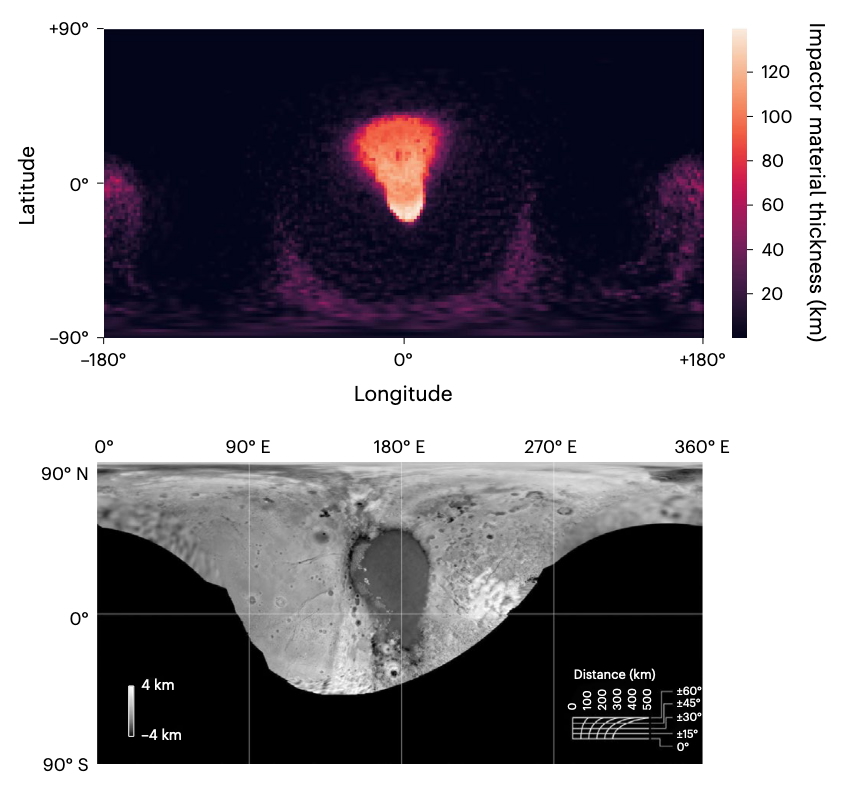} 
\caption{The distribution of impactor material after impact compared to Pluto’s observed elevation distribution.
Figure taken from \cite{Ballantyne2024}.}
\label{fig:pluto_heart}
\end{figure}

Pluto’s surface is dominated by Sputnik Planitia, a vast, pear-shaped basin that is widely believed to have formed through an impact event. However, existing models have struggled to explain its distinctive shape and its alignment near the Pluto-Charon axis. Ballantyne et al. \cite{Ballantyne2024} proposed an impact mechanism that accounts for both of these features using three-dimensional SPH simulations including material strength to model realistic collisions. The hypothesis does not require a cold, rigid crust overlying a liquid ocean. Instead, a scenario in which a differentiated ice-rock impactor, approximately \SI{730}{\kilo\meter} in diameter, collides at low velocity with a subsolidus Pluto-like target was considered. For a \SI{30}{\degree} collision, the simulations showed a new geologic region dominated by impactor material, namely a basin that closely reproduces the morphology of Sputnik Planitia (Fig.~\ref{fig:pluto_heart}), and a captured rocky impactor core that has penetrated the ice to accrete as a substantial, strength-supported mascon. This model provides an alternative explanation for the basin’s equatorial alignment and suggests that low-velocity collisions between trans-Neptunian objects occur in a regime in which strength effects can lead to impactor-dominated surface and subsurface structures.

\subsection{Small-body impacts and experiments}\label{sec:small_impacts}
\begin{figure}[ht]
\centering
\includegraphics[width=\textwidth]{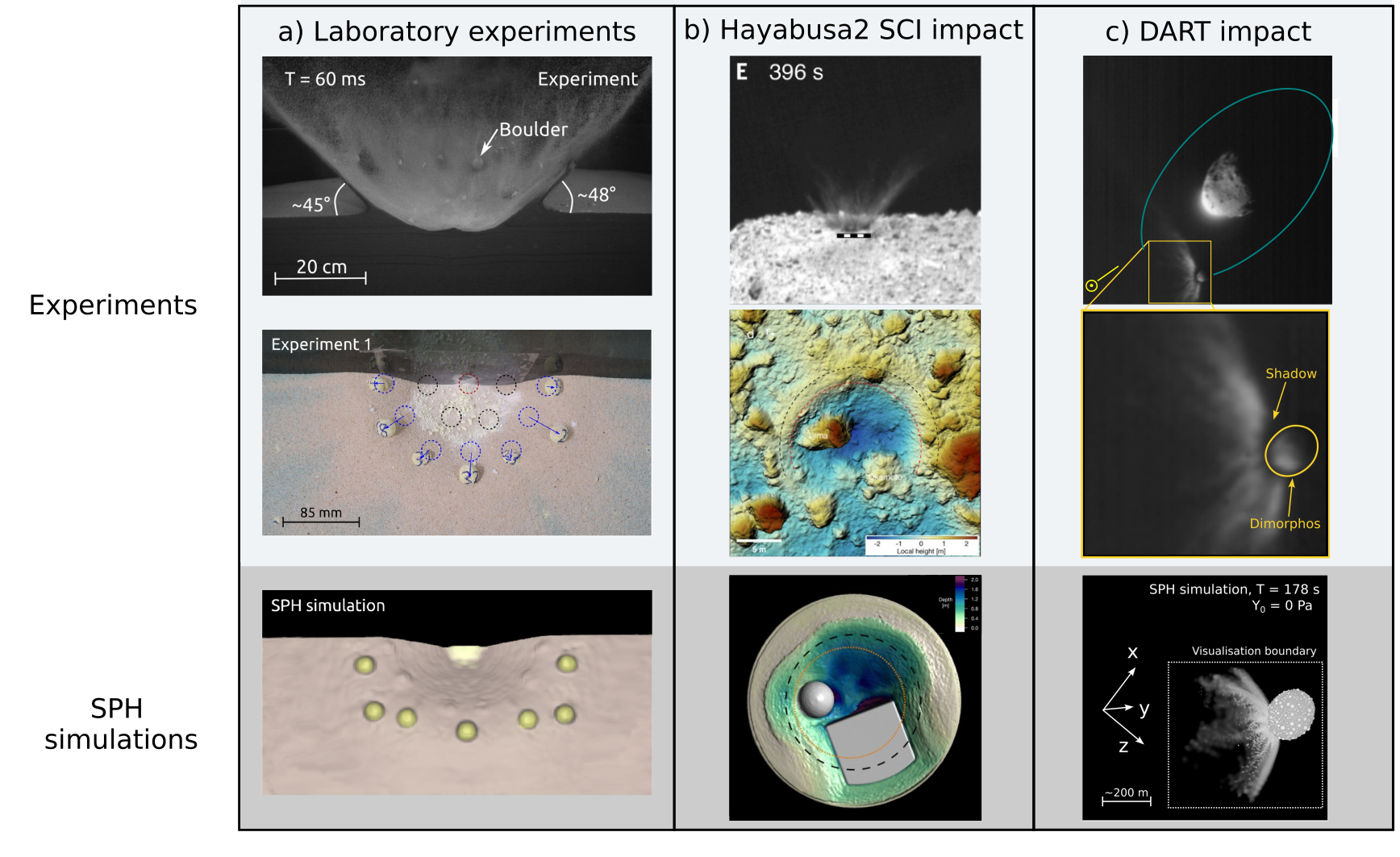} 
\caption{Laboratory (a) and space mission (b,c) experiments provide insights into the complex cratering processes on small bodies and serve as fundamental validation tests for numerical models.}
\label{fig:experiments}
\end{figure}

Small bodies, such as asteroids and comets, serve as natural laboratories for studying the conditions and mechanisms that shaped planetary bodies. Laboratory impact experiments have long been used to study impact phenomena and validate numerical models. To study conditions relevant for small asteroids, which are often considered to be rubble-pile objects, granular targets with embedded boulders are used to gain insights into the complex cratering processes on such bodies and to validate numerical models (Fig.~\ref{fig:experiments}a, \cite{Ormo2022}).

Recent space missions have provided unprecedented insights into impact processes on small bodies. One of the most significant breakthroughs was the Small Carry-on Impactor (SCI) experiment aboard JAXA’s Hayabusa2 mission at asteroid Ryugu, which in 2019 created the first fully documented artificial impact on an asteroid. This space impact experiment served as validation for shock physics codes (Fig.~\ref{fig:experiments} b). Using a hybrid scheme to model both the initial shock wave propagation and subsequent crater formation, the crater size and morphology were successfully reproduced \cite{Jutzi2022}. The simulations confirmed that the asteroid’s surface is extremely weak, with a cohesion of less than a few pascals.

NASA's DART mission provided another important advance in our understanding of impact processes on small bodies. The specific impact energy of the DART impact on asteroid Dimorphos was significantly larger than that of the SCI impact, leading to global-scale deformation of the target asteroid \cite{Raducan2022_global,raducanDART2024}. The comparison of impact modeling with the observations of the DART impact outcome (Fig.~\ref{fig:experiments} c) provided constraints on the structural and mechanical properties of Dimorphos, pointing towards a weak surface -- similar to what was found for the SCI impact on Ryugu. ESA's Hera mission, currently en route to the Didymos system, will perform a detailed characterization of the DART impact effects on Dimorphos and will allow further refinement of the impact models. The combined results from space mission experiments, laboratory investigations and numerical modeling, provide the underlying physics for studies of the collisional evolution of small body populations \cite{BottkeAsteroidsVI2015}.

\subsection{Low Velocity Mergers}

\begin{figure}[ht]
\centering
\includegraphics[width=\textwidth]{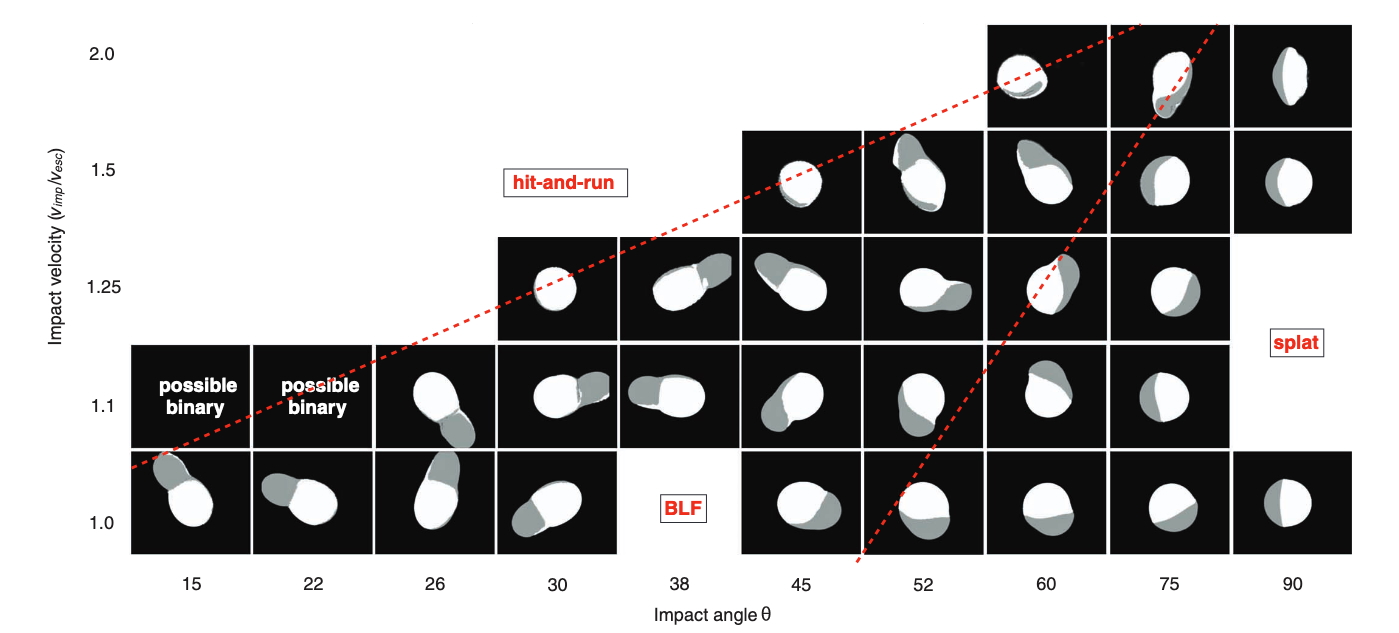} 
\caption{Outcomes from low-velocity mergers. A slice through the symmetry plane is shown; white is the target and gray is the impactor. Figure taken from \cite{jutzi_shape_2015}.}
\label{fig:low_vel_acc}
\end{figure}

The previous sections present various outcomes of high-velocity impact regimes. In the following, a very different regime is considered, with a focus on accretionary events at small scales. Typically, such collisions take place with velocities of the order of the mutual escape speed, which is below the speed of sound for scales smaller than a few \SI{1000}{\kilo\meter} (Fig.~\ref{fig:coll_regimes}). These "gentle" accretion events therefore do not involve strong shocks and conserve to some degree the structure of the accreted body; this can lead to interesting features like "splats" or bi-lobe shapes (\cite{jutzi_shape_2015}; Fig.~\ref{fig:low_vel_acc}). If the mergers take place in the tidal field of a planet, even more exotic structures can result \cite{leleu_peculiar_2018, wimarsson_rapid_2024}. Cometary nuclei, KBOs and small planetary moons imaged from flyby and rendezvous spacecraft show common evidence of such features. We note that such structures can also be a result of high-velocity (sub-)catastrophic impact events leading to the formation of two gravitationally bound clumps which eventually collide to form a bi-lobed shape (e.g. \cite{jutzi_formation_2017, jutzi_shape_2019}).

\section{Conclusions \& Outlook} \label{sec:outlook}
Recent advances in numerical impact modelling have profoundly enhanced our understanding of the formation and evolution of the Solar System. High-energy collisions between planetary bodies, known as giant impacts, are thought to play a central role in the final stages of planet formation, the origin of satellites, and the redistribution of mass and angular momentum. Smaller-scale impacts continue to shape the small body populations and planetary surfaces. Advances in SPH modelling have enabled the simulation of such processes at unprecedented levels of detail. Among these, advances of the Bern SPH code have enabled detailed end-to-end simulations of small body impacts, such as those investigated in the recent DART and SCI space mission experiments \cite{Jutzi2022, raducanDART2024}. The novel HPC code, \texttt{pkdgrav3} \cite{potterPKDGRAV3TrillionParticle2017, meierOriginJupitersFuzzy2025, Bussmann2025, meierSmoothedParticleHydrodynamicsinprep.}, stands out for its ability to model billions of particles, allowing one to resolve small structures such as a planet's crust, ocean or atmosphere or very low mass circumplanetary disks.

Ultra-high-resolution simulations have been enabled in recent years thanks to a shift from pure CPU systems to hybrid CPU-accelerator architectures, with most new high-performance computing (HPC) systems deriving the bulk of their performance from GPUs. This trend has spurred the development of GPU-enabled codes such as \texttt{pkdgrav3}. In contrast, legacy CPU-only codes struggle to secure compute time, as HPC centers prioritize projects that fully utilize modern hardware. The issue is compounded by GPU manufacturers focusing R\&D on AI-specific features (e.g. tensor cores, transformer engines), while traditional HPC-focused components receive less attention. To stay viable, simulation codes must adapt to leverage these evolving accelerator capabilities. 

State-of-the-art simulations that are restricted to the fluid regime are not always conclusive, as they lack the rheological complexity required to fully capture the behaviour of solid planetary materials under extreme stress and temperature conditions. Furthermore, time resolution remains a challenge: accurate modelling of impact scenarios requires bridging vastly different timescales, from the short-timescale initial high-velocity collision to the long-term dynamical evolution of resulting fragments. 

The limitations of earlier modelling approaches, often constrained by particle count or simplified physical assumptions, are becoming increasingly apparent in light of new observational efforts and space mission data. Combining state-of-the-art material models with highly efficient HPC codes such as \texttt{pkdgrav3} will enable the study of new collision scenarios, impact regimes, and outcomes at unprecedented levels of detail. For instance, billion-particle SPH simulations of DART-like impact scenarios will allow for fully resolving the impactor, boulder morphologies and distribution, and tracking the impact effects on the global asteroid -- including changes in surface morphology due to seismic waves and ejecta redistribution -- at sub-10 cm scales across the entire body. Observations of the DART impact outcome on asteroid Dimorphos by ESA's Hera mission will provide constraints for such advanced numerical models. Another example of a giant impact where both ultra-high resolution and physical realism are required is the formation of the Martian moons Phobos and Deimos \cite{rosenblatt_accretion_2016, canup_origin_2018, rosenblatt_formation_2020}. Billion-particle SPH simulations will accurately resolve the outer region of the very low mass disk where the moons form \cite{rosenblatt_accretion_2016} and provide detailed information on the predicted composition, thermal state and volatile depletion of the moon. These will play a key role for interpreting the samples returned by JAXA's Martian Moon eXplorer (MMX) expected in the early 2030s and solving the mistery of the origin of Phobos and Deimos.

PlanetS-NCCR encouraged collaboration between different groups and helped consolidate the efforts of impact modelling within Switzerland. Over the past decade, substantial contributions from Swiss research groups have driven major progress in the modelling of impact processes across different size regimes and physical scales. Nevertheless, the field remains dynamic and evolving. Future improvements in physical modelling, particularly with respect to material rheology, mixing, chemistry and phase transitions, and the coupling of impact codes with long-term interior and dynamical models will be essential. In parallel, new missions and remote-sensing observations will continue to provide increasingly accurate constraints, necessitating ongoing refinement of numerical models. The synergy between observational advances and high-resolution simulations promises to further deepen our understanding of planetary impacts, both within our Solar System and beyond.

\begin{acknowledgement}
This work has been carried out within the framework of the NCCR PlanetS supported by the Swiss National Science Foundation. 
CR, SDR and MJ acknowledge support from the SNSF (project 200021\_207359), 
CR acknowledges support from the SNSF (project 200021\_207359) and PlanetS (51NF40\_205606), Thomas Meier was supported by PlanetS (51NF40\_182901 and 51NF40\_205606).
\end{acknowledgement}

\ethics{Competing Interests}{The authors have no conflicts of interest to declare that are relevant to the content of this chapter.}

\bibliographystyle{spmpsci}
\bibliography{main}

@ARTICLE{Collins2014_dilatancy,
       author = {{Collins}, G.~S.},
        title = "{Numerical simulations of impact crater formation with dilatancy}",
      journal = {Journal of Geophysical Research (Planets)},
         year = 2014,
        month = dec,
       volume = {119},
       number = {12},
        pages = {2600-2619},
          doi = {10.1002/2014JE004708},
       adsurl = {https://ui.adsabs.harvard.edu/abs/2014JGRE..119.2600C}
}

@article{Canup2001a,
  title = {Origin of the {{Moon}} in a Giant Impact near the End of the {{Earth}}'s Formation},
  author = {Canup, Robin M. and Asphaug, Erik},
  year = {2001},
  month = aug,
  journal = {Nature},
  volume = {412},
  number = {6848},
  pages = {708--712},
  issn = {1476-4687},
  doi = {10.1038/35089010},
  urldate = {2018-09-06},
  copyright = {2001 Nature Publishing Group},
  langid = {english},
}

@article{Canup2004,
  title = {Simulations of a Late Lunar-Forming Impact},
  author = {Canup, Robin M.},
  year = {2004},
  month = apr,
  journal = {Icarus},
  volume = {168},
  number = {2},
  pages = {433--456},
  issn = {0019-1035},
  doi = {10.1016/j.icarus.2003.09.028},
  urldate = {2017-05-09},
}

@article{liuFormationJupitersDiluted2019,
  title = {The Formation of {{Jupiter}}'s Diluted Core by a Giant Impact},
  author = {Liu, Shang-Fei and Hori, Yasunori and M{\"u}ller, Simon and Zheng, Xiaochen and Helled, Ravit and Lin, Doug and Isella, Andrea},
  year = {2019},
  month = aug,
  journal = {Nature},
  volume = {572},
  number = {7769},
  pages = {355--357},
  issn = {1476-4687},
  doi = {10.1038/s41586-019-1470-2},
  urldate = {2019-08-19},
  copyright = {2019 The Author(s), under exclusive licence to Springer Nature Limited},
  langid = {english},
}

@article{benzOriginMercury2007,
  title = {The {{Origin}} of {{Mercury}}},
  author = {Benz, W. and Anic, A. and Horner, J. and Whitby, J. A.},
  year = {2007},
  month = oct,
  journal = {Space Science Reviews},
  volume = {132},
  number = {2-4},
  pages = {189--202},
  issn = {0038-6308, 1572-9672},
  doi = {10.1007/s11214-007-9284-1},
  urldate = {2017-04-30},
  langid = {english},
}

@article{Timpe2023,
  title = {A {{Systematic Survey}} of {{Moon-forming Giant Impacts}}. {{I}}. {{Nonrotating Bodies}}},
  author = {Timpe, Miles and Reinhardt, Christian and Meier, Thomas and Stadel, Joachim and Moore, Ben},
  year = {2023},
  month = dec,
  journal = {The Astrophysical Journal},
  volume = {959},
  number = {1},
  pages = {38},
  publisher = {The American Astronomical Society},
  issn = {0004-637X},
  doi = {10.3847/1538-4357/acfc40},
  urldate = {2024-01-20},
  langid = {english},
}

@article{Ballantyne2023,
  title = {Investigating the Feasibility of an Impact-Induced {{Martian Dichotomy}}},
  author = {Ballantyne, Harry A. and Jutzi, Martin and Golabek, Gregor J. and Mishra, Lokesh and Cheng, Kar Wai and Rozel, Antoine B. and Tackley, Paul J.},
  year = {2023},
  month = mar,
  journal = {Icarus},
  volume = {392},
  pages = {115395},
  issn = {0019-1035},
  doi = {10.1016/j.icarus.2022.115395},
  urldate = {2023-09-24},
}

@article{Ballantyne2024,
  title = {Sputnik {{Planitia}} as an Impactor Remnant Indicative of an Ancient Rocky Mascon in an Oceanless {{Pluto}}},
  author = {Ballantyne, Harry A. and Asphaug, Erik and Denton, C. Adeene and Emsenhuber, Alexandre and Jutzi, Martin},
  year = {2024},
  month = jun,
  journal = {Nature Astronomy},
  volume = {8},
  number = {6},
  pages = {748--755},
  publisher = {Nature Publishing Group},
  issn = {2397-3366},
  doi = {10.1038/s41550-024-02248-1},
  urldate = {2025-02-21},
  copyright = {2024 The Author(s)},
  langid = {english},
}

@article{Emsenhuber2018,
  title = {{{SPH}} Calculations of {{Mars-scale}} Collisions: {{The}} Role of the Equation of State, Material Rheologies, and Numerical Effects},
  shorttitle = {{{SPH}} Calculations of {{Mars-scale}} Collisions},
  author = {Emsenhuber, Alexandre and Jutzi, Martin and Benz, Willy},
  year = {2018},
  month = feb,
  journal = {Icarus},
  volume = {301},
  pages = {247--257},
  issn = {0019-1035},
  doi = {10.1016/j.icarus.2017.09.017},
  urldate = {2019-08-05},
}

@article{dengEnhancedMixingGiant2019,
  title = {Enhanced {{Mixing}} in {{Giant Impact Simulations}} with a {{New Lagrangian Method}}},
  author = {Deng, Hongping and Reinhardt, Christian and Benitez, Federico and Mayer, Lucio and Stadel, Joachim and Barr, Amy C.},
  date = {2019-01},
  journaltitle = {The Astrophysical Journal},
  shortjournal = {ApJ},
  journal = {The Astrophysical Journal},
  year = {2019},
  volume = {870},
  number = {2},
  pages = {127},
  issn = {0004-637X},
  doi = {10.3847/1538-4357/aaf399},
  url = {https://doi.org/10.3847%2F1538-4357%2Faaf399},
  urldate = {2019-01-25},
  langid = {english}
}

@article{dengPrimordialEarthMantle2019,
  title = {Primordial {{Earth Mantle Heterogeneity Caused}} by the {{Moon-forming Giant Impact}}?},
  author = {Deng, Hongping and Ballmer, Maxim D. and Reinhardt, Christian and Meier, Matthias M. M. and Mayer, Lucio and Stadel, Joachim and Benitez, Federico},
  year = {2019},
  month = dec,
  journal = {The Astrophysical Journal},
  volume = {887},
  number = {2},
  pages = {211},
  publisher = {American Astronomical Society},
  issn = {0004-637X},
  doi = {10.3847/1538-4357/ab50b9},
  url = {https://doi.org/10.3847%2F1538-4357%2Fab50b9},
  urldate = {2020-12-01},
  langid = {english}
}

@article{jutziModelingAsteroidCollisions2015,
  title = {Modeling Asteroid Collisions and Impact Processes},
  author = {Jutzi, Martin and Holsapple, Keith and W{\"u}nneman, Kai and Michel, Patrick},
  year = {2015},
  journal = {arXiv:1502.01844 [astro-ph]},
  eprint = {1502.01844},
  primaryclass = {astro-ph},
  doi = {10.2458/azu_uapress_9780816532131-ch035},
  url = {http://arxiv.org/abs/1502.01844},
  urldate = {2020-05-05},
  archiveprefix = {arXiv}
}

@article{meierEOSResolutionConspiracy2021,
  title = {The {{EOS}}/Resolution Conspiracy: Convergence in Proto-Planetary Collision Simulations},
  shorttitle = {The {{EOS}}/Resolution Conspiracy},
  author = {Meier, Thomas and Reinhardt, Christian and Stadel, Joachim G.},
  year = {2021},
  month = jun,
  journal = {Monthly Notices of the Royal Astronomical Society},
  volume = {505},
  number = {2},
  pages = {1806--1816},
  issn = {0035-8711, 1365-2966},
  doi = {10.1093/mnras/stab1441},
  url = {https://academic.oup.com/mnras/article/505/2/1806/6279686},
  urldate = {2021-09-06},
  langid = {english}
}

@unpublished{meierSmoothedParticleHydrodynamicsinprep.,
  title = {Smoothed {{Particle Hydrodynamics}} in Pkdgrav3 for Planetary Collision Simulations},
  author = {Meier, Thomas and Reinhardt, Christian and Potter, Douglas and Stadel, Joachim},
  year = {in prep.},
  note = {[Manuscript in preparation]}
}

@article{pierazzoValidationNumericalCodes2008,
  title = {Validation of Numerical Codes for Impact and Explosion Cratering: {{Impacts}} on Strengthless and Metal Targets},
  shorttitle = {Validation of Numerical Codes for Impact and Explosion Cratering},
  author = {Pierazzo, E. and Artemieva, N. and Asphaug, E. and Baldwin, E. C. and Cazamias, J. and Coker, R. and Collins, G. S. and Crawford, D. A. and Davison, T. and Elbeshausen, D. and Holsapple, K. A. and Housen, K. R. and Korycansky, D. G. and W{\"u}nnemann, K.},
  year = {2008},
  journal = {Meteoritics \& Planetary Science},
  volume = {43},
  number = {12},
  pages = {1917--1938},
  issn = {1945-5100},
  doi = {10.1111/j.1945-5100.2008.tb00653.x},
  url = {https://onlinelibrary.wiley.com/doi/abs/10.1111/j.1945-5100.2008.tb00653.x},
  urldate = {2025-01-07},
  copyright = {2008 The Meteoritical Society},
  langid = {english}
}

@article{potterPKDGRAV3TrillionParticle2017,
  title = {{{PKDGRAV3}}: Beyond Trillion Particle Cosmological Simulations for the next Era of Galaxy Surveys},
  shorttitle = {{{PKDGRAV3}}},
  author = {Potter, Douglas and Stadel, Joachim and Teyssier, Romain},
  year = {2017},
  month = dec,
  journal = {Computational Astrophysics and Cosmology},
  volume = {4},
  number = {1},
  pages = {2},
  issn = {2197-7909},
  doi = {10.1186/s40668-017-0021-1},
  url = {https://comp-astrophys-cosmol.springeropen.com/articles/10.1186/s40668-017-0021-1},
  urldate = {2021-03-15},
  langid = {english}
}

@article{priceSmoothedParticleHydrodynamics2012,
  title = {Smoothed Particle Hydrodynamics and Magnetohydrodynamics},
  author = {Price, Daniel J.},
  year = {2012},
  month = feb,
  journal = {Journal of Computational Physics},
  series = {Special {{Issue}}: {{Computational Plasma Physics}}},
  volume = {231},
  number = {3},
  pages = {759--794},
  issn = {0021-9991},
  doi = {10.1016/j.jcp.2010.12.011},
  url = {https://www.sciencedirect.com/science/article/pii/S0021999110006753},
  urldate = {2021-03-09},
  langid = {english}
}

@INCOLLECTION{BottkeAsteroidsVI2015,
       author = {{Bottke}, W.~F. and {Bro{\v{z}}}, M. and {O'Brien}, D.~P. and {Campo Bagatin}, A. and {Morbidelli}, A. and {Marchi}, S.},
        title = "{The Collisional Evolution of the Main Asteroid Belt}",
    booktitle = {Asteroids IV},
         year = 2015,
       editor = {{Michel}, Patrick and {DeMeo}, Francesca E. and {Bottke}, William F.},
    publisher = {University of Arizona Press},
        pages = {701-724},
          doi = {10.2458/azu_uapress_9780816532131-ch036},
       adsurl = {https://ui.adsabs.harvard.edu/abs/2015aste.book..701B}
}

@ARTICLE{raducanDART2024,
       author = {{Raducan}, S.~D. and {Jutzi}, M. and {Cheng}, A.~F. and {Zhang}, Y. and {Barnouin}, O. and {Collins}, G.~S. and {Daly}, R.~T. and {Davison}, T.~M. and {Ernst}, C.~M. and {Farnham}, T.~L. and {Ferrari}, F. and {Hirabayashi}, M. and {Kumamoto}, K.~M. and {Michel}, P. and {Murdoch}, N. and {Nakano}, R. and {Pajola}, M. and {Rossi}, A. and {Agrusa}, H.~F. and {Barbee}, B.~W. and {Syal}, M. Bruck and {Chabot}, N.~L. and {Dotto}, E. and {Fahnestock}, E.~G. and {Hasselmann}, P.~H. and {Herreros}, I. and {Ivanovski}, S. and {Li}, J. -Y. and {Lucchetti}, A. and {Luther}, R. and {Orm{\"o}}, J. and {Owen}, M. and {Pravec}, P. and {Rivkin}, A.~S. and {Robin}, C.~Q. and {S{\'a}nchez}, P. and {Tusberti}, F. and {W{\"u}nnemann}, K. and {Zinzi}, A. and {Epifani}, E. Mazzotta and {Manzoni}, C. and {May}, B.~H.},
        title = "{Physical properties of asteroid Dimorphos as derived from the DART impact}",
      journal = {Nature Astronomy},
         year = 2024,
        month = apr,
       volume = {8},
        pages = {445-455},
          doi = {10.1038/s41550-024-02200-3},
archivePrefix = {arXiv},
       eprint = {2403.00667},
 primaryClass = {astro-ph.EP},
       adsurl = {https://ui.adsabs.harvard.edu/abs/2024NatAs...8..445R}
}

@article{reinhardtBifurcationHistoryUranus2020,
  title = {Bifurcation in the History of {{Uranus}} and {{Neptune}}: The Role of Giant Impacts},
  shorttitle = {Bifurcation in the History of {{Uranus}} and {{Neptune}}},
  author = {Reinhardt, Christian and Chau, Alice and Stadel, Joachim and Helled, Ravit},
  year = {2020},
  month = mar,
  journal = {Monthly Notices of the Royal Astronomical Society},
  volume = {492},
  number = {4},
  pages = {5336--5353},
  publisher = {Oxford Academic},
  issn = {0035-8711},
  doi = {10.1093/mnras/stz3271},
  url = {https://academic.oup.com/mnras/article/492/4/5336/5637902},
  urldate = {2020-08-10},
  langid = {english}
}

@article{reinhardtNumericalAspectsGiant2017,
  title = {Numerical Aspects of {{Giant Impact}} Simulations},
  author = {Reinhardt, Christian and Stadel, Joachim},
  year = {2017},
  month = jan,
  journal = {Monthly Notices of the Royal Astronomical Society},
  volume = {467},
  doi = {10.1093/mnras/stx322},
  url = {http://adsabs.harvard.edu/abs/2017MNRAS.467.4252R}
}

@article{ruiz-bonillaDealingDensityDiscontinuities2022,
  title = {Dealing with Density Discontinuities in Planetary {{SPH}} Simulations},
  author = {{Ruiz-Bonilla}, S and Borrow, J and Eke, V R and Kegerreis, J A and Massey, R J and Sandnes, T D and Teodoro, L F A},
  year = {2022},
  month = may,
  journal = {Monthly Notices of the Royal Astronomical Society},
  volume = {512},
  number = {3},
  pages = {4660--4668},
  issn = {0035-8711},
  doi = {10.1093/mnras/stac857},
  url = {https://doi.org/10.1093/mnras/stac857},
  urldate = {2023-10-25}
}

@article{Collins2004,
	title = {Modeling damage and deformation in impact simulations},
	volume = {39},
	issn = {1945-5100},
	doi = {10.1111/j.1945-5100.2004.tb00337.x},
	number = {2},
	journal = {Meteorit. Planet. Sci},
	author = {{Collins}, Gareth S. and Melosh, H. Jay and Ivanov, Boris A.},
	year = {2004},
	pages = {217--231},
}

@article{Jutzi2008,
	title = {Numerical simulations of impacts involving porous bodies: {I}. {Implementing} sub-resolution porosity in a {3D} {SPH} {Hydrocode}},
	volume = {198},
	shorttitle = {Numerical simulations of impacts involving porous bodies},
	doi = {10.1016/j.icarus.2008.06.013},
	number = {1},
	journal = {Icarus},
	author = {Jutzi, M. and Benz, W. and Michel, P.},
	month = nov,
	year = {2008},
	pages = {242--255},
}

@article{Jutzi2009,
	title = {Numerical simulations of impacts involving porous bodies: {II}. Comparison with laboratory experiments},
	volume = {201},
	doi = {10.1016/j.icarus.2009.01.018},
	shorttitle = {Numerical simulations of impacts involving porous bodies},
	pages = {802--813},
	number = {2},
	journal = {Icarus},
	author = {Jutzi, Martin and Michel, Patrick and Hiraoka, Kensuke and Nakamura, Akiko M. and Benz, Willy},
	year = {2009},
}

@article{Benz1994,
	title = {Impact {Simulations} with {Fracture}. {I}. {Method} and {Tests}},
	volume = {107},
	issn = {0019-1035},
	doi = {10.1006/icar.1994.1009},
	number = {1},
	journal = {Icarus},
	author = {Benz, W. and Asphaug, E.},
	year = {1994},
	pages = {98--116},
}

@article{Benz1995,
	series = {Particle {Simulation} {Methods}},
	title = {Simulations of brittle solids using smooth particle hydrodynamics},
	volume = {87},
	issn = {0010-4655},
	number = {1},
	journal = {Comput. Phys. Commun},
	author = {Benz, W. and Asphaug, E.},
	year = {1995},
	pages = {253--265},
}

@article{Jutzi2015,
	title = {{SPH} calculations of asteroid disruptions: {The} role of pressure dependent failure models},
	volume = {107},
	issn = {0032-0633},
	doi = {10.1016/j.pss.2014.09.012},
	journal = {Planet. Space Sci},
	author = {Jutzi, Martin},
	year = {2015},
	pages = {3--9},
}

@article{Jutzi2022,
	title = {Constraining surface properties of asteroid (162173) Ryugu from numerical simulations of Hayabusa2 mission impact experiment},
	volume = {13},
	rights = {2022 The Author(s)},
	issn = {2041-1723},
	doi = {10.1038/s41467-022-34540-x},
	pages = {7134},
	number = {1},
	journal = {Nature Communications},
	author = {Jutzi, Martin and Raducan, Sabina D. and Zhang, Yun and Michel, Patrick and Arakawa, Masahiko},
	year = {2022},

}

@article{Raducan2022_global,
	title = {Global-scale Reshaping and Resurfacing of Asteroids by Small-scale Impacts, with Applications to the {DART} and Hera Missions},
	volume = {3},
	issn = {2632-3338},
	doi = {10.3847/PSJ/ac67a7},
	pages = {128},
	number = {6},
	journal = {Planetary Science Journal},
	author = {Raducan, Sabina D. and Jutzi, Martin},
	year = {2022},
}

@article{Raducan2024_lessons,
	title = {Lessons {Learned} from {NASA}’s {DART} {Impact} about {Disrupting} {Rubble}-pile {Asteroids}},
	volume = {5},
	doi = {10.3847/PSJ/ad29f6},
	number = {3},
	journal = {The Planetary Science Journal},
	author = {Raducan, S. D. and Jutzi, M. and Merrill, C. C. and Michel, P. and Zhang, Y. and Hirabayashi, M. and Mainzer, A.},
	year = {2024},
	pages = {79},
}

@article{Ormo2022,
	title = {Boulder exhumation and segregation by impacts on rubble-pile asteroids},
	volume = {594},
	issn = {0012-821X},
	doi = {10.1016/j.epsl.2022.117713},
	pages = {117713},
	journal = {Earth and Planetary Science Letters},
	author = {Ormö, J. and Raducan, S. D. and Jutzi, M. and Herreros, M. I. and Luther, R. and Collins, G. S. and Wünnemann, K. and Mora-Rueda, M. and Hamann, C.},
	year = {2022},
}

@article{Cameron1976,
  title = {The {{Origin}} of the {{Moon}}},
  author = {Cameron, A. G. W. and Ward, W. R.},
  year = {1976},
  month = mar,
  journal = {Lunar and Planetary Science Conference},
  volume = {7},
  pages = {120},
  urldate = {2019-08-19},
  langid = {english},
}

@article{Hartmann1975,
  title = {Satellite-Sized Planetesimals and Lunar Origin},
  author = {Hartmann, William K. and Davis, Donald R.},
  year = {1975},
  month = apr,
  journal = {Icarus},
  volume = {24},
  number = {4},
  pages = {504--515},
  issn = {0019-1035},
  doi = {10.1016/0019-1035(75)90070-6},
  urldate = {2025-02-21},
}

@article{meier_systematic_2024,
    title = {A {Systematic} {Survey} of {Moon}-forming {Giant} {Impacts}. {II}. {Rotating} {Bodies}},
    volume = {978},
    issn = {0004-637X},
    url = {https://dx.doi.org/10.3847/1538-4357/ad9248},
    doi = {10.3847/1538-4357/ad9248},
    language = {en},
    number = {1},
    urldate = {2025-01-09},
    journal = {The Astrophysical Journal},
    author = {Meier, Thomas and Reinhardt, Christian and Timpe, Miles and Stadel, Joachim and Moore, Ben},
    month = dec,
    year = {2024},
    note = {Publisher: The American Astronomical Society},
    pages = {11},
}

@article{hyodo_impact_2017,
    title = {On the {Impact} {Origin} of {Phobos} and {Deimos}. {I}. {Thermodynamic} and {Physical} {Aspects}},
    volume = {845},
    issn = {0004-637X},
    url = {http://stacks.iop.org/0004-637X/845/i=2/a=125},
    doi = {10.3847/1538-4357/aa81c4},
    language = {en},
    number = {2},
    urldate = {2017-12-03},
    journal = {The Astrophysical Journal},
    author = {Hyodo, Ryuki and Genda, Hidenori and Charnoz, Sébastien and Rosenblatt, Pascal},
    year = {2017},
    pages = {125},
}

@article{woo_did_2022,
    title = {Did {Uranus}' regular moons form via a rocky giant impactor?},
    volume = {375},
    issn = {0019-1035},
    url = {https://www.sciencedirect.com/science/article/pii/S0019103521004851},
    doi = {10.1016/j.icarus.2021.114842},
    urldate = {2023-09-24},
    journal = {Icarus},
    author = {Woo, Jason Man Yin and Reinhardt, Christian and Cilibrasi, Marco and Chau, Alice and Helled, Ravit and Stadel, Joachim},
    month = mar,
    year = {2022},
    pages = {114842},
}

@article{chau_could_2021,
    title = {Could {Uranus} and {Neptune} form by collisions of planetary embryos?},
    volume = {502},
    issn = {0035-8711},
    url = {https://doi.org/10.1093/mnras/staa4021},
    doi = {10.1093/mnras/staa4021},
    number = {2},
    urldate = {2025-01-09},
    journal = {Monthly Notices of the Royal Astronomical Society},
    author = {Chau, Alice and Reinhardt, Christian and Izidoro, André and Stadel, Joachim and Helled, Ravit},
    month = apr,
    year = {2021},
    pages = {1647--1660},
}

@article{canup_forming_2012,
    title = {Forming a {Moon} with an {Earth}-like {Composition} via a {Giant} {Impact}},
    volume = {338},
    url = {https://ui.adsabs.harvard.edu/abs/2012Sci...338.1052C/abstract},
    doi = {10.1126/science.1226073},
    language = {en},
    number = {6110},
    urldate = {2019-08-19},
    journal = {Science},
    author = {Canup, Robin M.},
    month = nov,
    year = {2012},
    pages = {1052},
}

@article{cuk_making_2012,
    title = {Making the {Moon} from a {Fast}-{Spinning} {Earth}: {A} {Giant} {Impact} {Followed} by {Resonant} {Despinning}},
    volume = {338},
    shorttitle = {Making the {Moon} from a {Fast}-{Spinning} {Earth}},
    url = {https://ui.adsabs.harvard.edu/abs/2012Sci...338.1047C/abstract},
    doi = {10.1126/science.1225542},
    language = {en},
    number = {6110},
    urldate = {2019-08-19},
    journal = {Science},
    author = {Ćuk, Matija and Stewart, Sarah T.},
    month = nov,
    year = {2012},
    pages = {1047},
}

@article{asphaug_mercury_2014,
    title = {Mercury and other iron-rich planetary bodies as relics of inefficient accretion},
    volume = {7},
    copyright = {© 2014 Nature Publishing Group},
    issn = {1752-0894},
    url = {http://www.nature.com/ngeo/journal/v7/n8/full/ngeo2189.html},
    doi = {10.1038/ngeo2189},
    language = {en},
    number = {8},
    urldate = {2017-02-01},
    journal = {Nature Geoscience},
    author = {Asphaug, E. and Reufer, A.},
    month = aug,
    year = {2014},
    pages = {564--568},
}

@article{chau_forming_2018,
    title = {Forming {Mercury} by {Giant} {Impacts}},
    volume = {865},
    issn = {0004-637X},
    url = {http://stacks.iop.org/0004-637X/865/i=1/a=35},
    doi = {10.3847/1538-4357/aad8b0},
    language = {en},
    number = {1},
    urldate = {2018-11-26},
    journal = {The Astrophysical Journal},
    author = {Chau, Alice and Reinhardt, Christian and Helled, Ravit and Stadel, Joachim},
    year = {2018},
    pages = {35},
}

@article{stewart_shock_2020,
    title = {The shock physics of giant impacts: {Key} requirements for the equations of state},
    volume = {2272},
    issn = {0094-243X},
    shorttitle = {The shock physics of giant impacts},
    url = {https://aip.scitation.org/doi/abs/10.1063/12.0000946},
    doi = {10.1063/12.0000946},
    number = {1},
    urldate = {2022-02-14},
    journal = {AIP Conference Proceedings},
    author = {Stewart, Sarah and Davies, Erik and Duncan, Megan and Lock, Simon and Root, Seth and Townsend, Joshua and Kraus, Richard and Caracas, Razvan and Jacobsen, Stein},
    month = nov,
    year = {2020},
    note = {Publisher: American Institute of Physics},
    pages = {080003},
}

@article{quintana_frequency_2016,
    title = {The {Frequency} of {Giant} {Impacts} on {Earth}-like {Worlds}},
    volume = {821},
    issn = {0004-637X},
    url = {http://stacks.iop.org/0004-637X/821/i=2/a=126},
    doi = {10.3847/0004-637X/821/2/126},
    language = {en},
    number = {2},
    urldate = {2017-02-01},
    journal = {The Astrophysical Journal},
    author = {Quintana, Elisa V. and Barclay, Thomas and Borucki, William J. and Rowe, Jason F. and Chambers, John E.},
    year = {2016},
    pages = {126},
}

@article{agnor_character_1999,
    title = {On the {Character} and {Consequences} of {Large} {Impacts} in the {Late} {Stage} of {Terrestrial} {Planet} {Formation}},
    volume = {142},
    issn = {0019-1035},
    url = {http://www.sciencedirect.com/science/article/pii/S0019103599962012},
    doi = {10.1006/icar.1999.6201},
    number = {1},
    urldate = {2019-08-19},
    journal = {Icarus},
    author = {Agnor, Craig B. and Canup, Robin M. and Levison, Harold F.},
    month = nov,
    year = {1999},
    pages = {219--237},
}

@article{chambers_making_2001,
    title = {Making {More} {Terrestrial} {Planets}},
    volume = {152},
    issn = {0019-1035},
    url = {http://www.sciencedirect.com/science/article/pii/S0019103501966394},
    doi = {10.1006/icar.2001.6639},
    language = {en},
    number = {2},
    urldate = {2021-01-24},
    journal = {Icarus},
    author = {Chambers, J. E.},
    month = aug,
    year = {2001},
    pages = {205--224},
}

@article{izidoro_accretion_2015,
    title = {Accretion of {Uranus} and {Neptune} from inward-migrating planetary embryos blocked by {Jupiter} and {Saturn}},
    volume = {582},
    copyright = {© ESO, 2015},
    issn = {0004-6361, 1432-0746},
    url = {http://dx.doi.org/10.1051/0004-6361/201425525},
    doi = {10.1051/0004-6361/201425525},
    language = {en},
    urldate = {2017-06-22},
    journal = {Astronomy \& Astrophysics},
    author = {Izidoro, André and Morbidelli, Alessandro and Raymond, Sean N. and Hersant, Franck and Pierens, Arnaud},
    month = oct,
    year = {2015},
    pages = {A99},
}

@article{wurm_photophoretic_2013,
    title = {{PHOTOPHORETIC} {SEPARATION} {OF} {METALS} {AND} {SILICATES}: {THE} {FORMATION} {OF} {MERCURY}-{LIKE} {PLANETS} {AND} {METAL} {DEPLETION} {IN} {CHONDRITES}},
    volume = {769},
    issn = {0004-637X},
    shorttitle = {{PHOTOPHORETIC} {SEPARATION} {OF} {METALS} {AND} {SILICATES}},
    url = {https://doi.org/10.1088%2F0004-637x%2F769%2F1%2F78},
    doi = {10.1088/0004-637X/769/1/78},
    language = {en},
    number = {1},
    urldate = {2020-04-04},
    journal = {The Astrophysical Journal},
    author = {Wurm, Gerhard and Trieloff, Mario and Rauer, Heike},
    month = may,
    year = {2013},
    note = {Publisher: IOP Publishing},
    pages = {78},
}

@article{lewis_metalsilicate_1972,
    title = {Metal/silicate fractionation in the solar system},
    volume = {15},
    issn = {0012-821X},
    url = {http://www.sciencedirect.com/science/article/pii/0012821X72901744},
    doi = {10.1016/0012-821X(72)90174-4},
    language = {en},
    number = {3},
    urldate = {2020-04-04},
    journal = {Earth and Planetary Science Letters},
    author = {Lewis, John S.},
    month = jul,
    year = {1972},
    pages = {286--290},
}

@article{aguichine_rocklines_2020,
    title = {Rocklines as {Cradles} for {Refractory} {Solids} in the {Protosolar} {Nebula}},
    volume = {901},
    issn = {0004-637X},
    url = {https://doi.org/10.3847/1538-4357/abaf47},
    doi = {10.3847/1538-4357/abaf47},
    language = {en},
    number = {2},
    urldate = {2022-02-28},
    journal = {The Astrophysical Journal},
    author = {Aguichine, Artyom and Mousis, Olivier and Devouard, Bertrand and Ronnet, Thomas},
    month = sep,
    year = {2020},
    note = {Publisher: American Astronomical Society},
    pages = {97},
}

@article{wetherill_occurrence_1985,
    title = {Occurrence of {Giant} {Impacts} {During} the {Growth} of the {Terrestrial} {Planets}},
    volume = {228},
    copyright = {1985 by the American Association for the Advancement of Science.},
    issn = {0036-8075, 1095-9203},
    url = {https://science.sciencemag.org/content/228/4701/877},
    doi = {10.1126/science.228.4701.877},
    language = {en},
    number = {4701},
    urldate = {2019-08-19},
    journal = {Science},
    author = {Wetherill, George W.},
    month = may,
    year = {1985},
    pmid = {17815054},
    pages = {877--879},
}

@article{valletta_possible_2022,
    title = {Possible {In} {Situ} {Formation} of {Uranus} and {Neptune} via {Pebble} {Accretion}},
    volume = {931},
    issn = {0004-637X},
    url = {https://dx.doi.org/10.3847/1538-4357/ac5f52},
    doi = {10.3847/1538-4357/ac5f52},
    language = {en},
    number = {1},
    urldate = {2025-03-04},
    journal = {The Astrophysical Journal},
    author = {Valletta, Claudio and Helled, Ravit},
    month = may,
    year = {2022},
    note = {Publisher: The American Astronomical Society},
    pages = {21},
}

@article{eriksson_can_2023,
    title = {Can {Uranus} and {Neptune} form concurrently via pebble, gas, and planetesimal accretion?},
    volume = {526},
    issn = {0035-8711},
    url = {https://doi.org/10.1093/mnras/stad3007},
    doi = {10.1093/mnras/stad3007},
    number = {4},
    urldate = {2025-03-04},
    journal = {Monthly Notices of the Royal Astronomical Society},
    author = {Eriksson, Linn E J and Mol Lous, Marit A S and Shibata, Sho and Helled, Ravit},
    month = dec,
    year = {2023},
    pages = {4860--4876},
}

@article{esteves_accretion_2025,
    title = {Accretion of {Uranus} and {Neptune}: {Confronting} different giant impact scenarios},
    volume = {429},
    issn = {0019-1035},
    shorttitle = {Accretion of {Uranus} and {Neptune}},
    url = {https://www.sciencedirect.com/science/article/pii/S0019103524004883},
    doi = {10.1016/j.icarus.2024.116428},
    urldate = {2025-03-04},
    journal = {Icarus},
    author = {Esteves, Leandro and Izidoro, André and Winter, Othon C.},
    month = mar,
    year = {2025},
    pages = {116428},
}

@inproceedings{stevensonUranusNeptuneDichotomyRole1986,
  title = {The {{Uranus-Neptune Dichotomy}}: The {{Role}} of {{Giant Impacts}}},
  shorttitle = {The {{Uranus-Neptune Dichotomy}}},
  booktitle = {Lunar and {{Planetary Science Conference}}},
  author = {Stevenson, D. J.},
  year = {1986},
  month = mar,
  pages = {1011--1012},
  url = {https://ui.adsabs.harvard.edu/abs/1986LPI....17.1011S},
  urldate = {2025-04-25}
}

@article{podolak_what_2012,
    title = {What {Do} {We} {Really} {Know} about {Uranus} and {Neptune}?},
    volume = {759},
    issn = {2041-8205},
    url = {http://stacks.iop.org/2041-8205/759/i=2/a=L32},
    doi = {10.1088/2041-8205/759/2/L32},
    language = {en},
    number = {2},
    urldate = {2017-05-03},
    journal = {The Astrophysical Journal Letters},
    author = {Podolak, M. and Helled, R.},
    year = {2012},
    pages = {L32},
}

@article{cheng_combined_2024,
    title = {Combined impact and interior evolution models in three dimensions indicate a southern impact origin of the {Martian} {Dichotomy}},
    volume = {420},
    issn = {0019-1035},
    url = {https://www.sciencedirect.com/science/article/pii/S0019103524001970},
    doi = {10.1016/j.icarus.2024.116137},
    urldate = {2025-02-26},
    journal = {Icarus},
    author = {Cheng, Kar Wai and Ballantyne, Harry A. and Golabek, Gregor J. and Jutzi, Martin and Rozel, Antoine B. and Tackley, Paul J.},
    month = sep,
    year = {2024},
    pages = {116137},
}

@article{benz_origin_1989,
    title = {The origin of the {Moon} and the single-impact hypothesis {III}},
    volume = {81},
    issn = {0019-1035},
    url = {http://www.sciencedirect.com/science/article/pii/0019103589901292},
    doi = {10.1016/0019-1035(89)90129-2},
    number = {1},
    urldate = {2018-05-14},
    journal = {Icarus},
    author = {Benz, W. and Cameron, A. G. W. and Melosh, H. J.},
    month = sep,
    year = {1989},
    pages = {113--131},
}

@article{rosenblatt_accretion_2016,
    title = {Accretion of {Phobos} and {Deimos} in an extended debris disc stirred by transient moons},
    volume = {9},
    copyright = {2016 Springer Nature Limited},
    issn = {1752-0908},
    url = {https://www.nature.com/articles/ngeo2742},
    doi = {10.1038/ngeo2742},
    language = {en},
    number = {8},
    urldate = {2024-10-07},
    journal = {Nature Geoscience},
    author = {Rosenblatt, Pascal and Charnoz, Sébastien and Dunseath, Kevin M. and Terao-Dunseath, Mariko and Trinh, Antony and Hyodo, Ryuki and Genda, Hidenori and Toupin, Stéven},
    month = aug,
    year = {2016},
    note = {Publisher: Nature Publishing Group},
    pages = {581--583},
}

@article{asphaug_global_2015,
    title = {Global {Scale} {Impacts}},
    url = {http://arxiv.org/abs/1504.02389},
    doi = {10.2458/azu_uapress_9780816532131-ch034},
    urldate = {2017-11-30},
    journal = {arXiv:1504.02389 [astro-ph]},
    author = {Asphaug, Erik and Collins, Gareth and Jutzi, Martin},
    year = {2015},
    note = {arXiv: 1504.02389},
}

@article{wiechert_oxygen_2001,
    title = {Oxygen {Isotopes} and the {Moon}-{Forming} {Giant} {Impact}},
    volume = {294},
    issn = {0036-8075, 1095-9203},
    url = {https://science.sciencemag.org/content/294/5541/345},
    doi = {10.1126/science.1063037},
    language = {en},
    number = {5541},
    urldate = {2020-04-04},
    journal = {Science},
    author = {Wiechert, U. and Halliday, A. N. and Lee, D.-C. and Snyder, G. A. and Taylor, L. A. and Rumble, D.},
    month = oct,
    year = {2001},
    pmid = {11598294},
    note = {Publisher: American Association for the Advancement of Science
Section: Report},
    pages = {345--348},
}

@article{zhang_proto-earth_2012,
    title = {The proto-{Earth} as a significant source of lunar material},
    volume = {5},
    copyright = {2012 Nature Publishing Group},
    issn = {1752-0908},
    url = {https://www.nature.com/articles/ngeo1429},
    doi = {10.1038/ngeo1429},
    language = {en},
    number = {4},
    urldate = {2020-04-04},
    journal = {Nature Geoscience},
    author = {Zhang, Junjun and Dauphas, Nicolas and Davis, Andrew M. and Leya, Ingo and Fedkin, Alexei},
    month = apr,
    year = {2012},
    note = {Number: 4
Publisher: Nature Publishing Group},
    pages = {251--255},
}

@article{touboul_late_2007,
    title = {Late formation and prolonged differentiation of the {Moon} inferred from {W} isotopes in lunar metals},
    volume = {450},
    copyright = {2007 Springer Nature Limited},
    issn = {1476-4687},
    url = {https://www.nature.com/articles/nature06428},
    doi = {10.1038/nature06428},
    language = {en},
    number = {7173},
    urldate = {2025-03-05},
    journal = {Nature},
    author = {Touboul, M. and Kleine, T. and Bourdon, B. and Palme, H. and Wieler, R.},
    month = dec,
    year = {2007},
    note = {Publisher: Nature Publishing Group},
    pages = {1206--1209},
}

@article{bottke_fossilized_2005,
    title = {The fossilized size distribution of the main asteroid belt},
    volume = {175},
    issn = {0019-1035},
    url = {https://www.sciencedirect.com/science/article/pii/S0019103504003811},
    doi = {10.1016/j.icarus.2004.10.026},
    number = {1},
    urldate = {2025-03-05},
    journal = {Icarus},
    author = {Bottke, William F. and Durda, Daniel D. and Nesvorný, David and Jedicke, Robert and Morbidelli, Alessandro and Vokrouhlický, David and Levison, Hal},
    month = may,
    year = {2005},
    pages = {111--140},
}

@article{morbidelli_asteroids_2009,
    title = {Asteroids were born big},
    volume = {204},
    issn = {0019-1035},
    url = {https://www.sciencedirect.com/science/article/pii/S0019103509003029},
    doi = {10.1016/j.icarus.2009.07.011},
    number = {2},
    urldate = {2025-03-05},
    journal = {Icarus},
    author = {Morbidelli, Alessandro and Bottke, William F. and Nesvorný, David and Levison, Harold F.},
    month = dec,
    year = {2009},
    pages = {558--573},
}

@inproceedings{farinellaAsteroidFamiliesOld1996,
  title = {Asteroid {{Families}}, {{Old}} and {{Young}}},
  booktitle = {Completing the {{Inventory}} of the {{Solar System}}},
  author = {Farinella, P. and Davis, D. R. and Marzari, F.},
  year = {1996},
  month = jan,
  volume = {107},
  pages = {45--55},
  url = {https://ui.adsabs.harvard.edu/abs/1996ASPC..107...45F},
  urldate = {2025-04-25}
}

@article{fujiwara_rubble-pile_2006,
    title = {The {Rubble}-{Pile} {Asteroid} {Itokawa} as {Observed} by {Hayabusa}},
    volume = {312},
    url = {https://www.science.org/doi/10.1126/science.1125841},
    doi = {10.1126/science.1125841},
    number = {5778},
    urldate = {2025-03-05},
    journal = {Science},
    author = {Fujiwara, A. and Kawaguchi, J. and Yeomans, D. K. and Abe, M. and Mukai, T. and Okada, T. and Saito, J. and Yano, H. and Yoshikawa, M. and Scheeres, D. J. and Barnouin-Jha, O. and Cheng, A. F. and Demura, H. and Gaskell, R. W. and Hirata, N. and Ikeda, H. and Kominato, T. and Miyamoto, H. and Nakamura, A. M. and Nakamura, R. and Sasaki, S. and Uesugi, K.},
    month = jun,
    year = {2006},
    note = {Publisher: American Association for the Advancement of Science},
    pages = {1330--1334},
}

@article{kaib_feeding_2015,
    title = {The feeding zones of terrestrial planets and insights into {Moon} formation},
    volume = {252},
    issn = {0019-1035},
    url = {https://www.sciencedirect.com/science/article/pii/S0019103515000196},
    doi = {10.1016/j.icarus.2015.01.013},
    language = {en},
    urldate = {2022-11-17},
    journal = {Icarus},
    author = {Kaib, Nathan A. and Cowan, Nicolas B.},
    month = may,
    year = {2015},
    pages = {161--174},
}

@article{mastrobuono-battisti_composition_2017,
    title = {The composition of {Solar} system asteroids and {Earth}/{Mars} moons, and the {Earth}–{Moon} composition similarity},
    volume = {469},
    issn = {0035-8711},
    url = {https://academic.oup.com/mnras/article/469/3/3597/3795552},
    doi = {10.1093/mnras/stx1054},
    language = {en},
    number = {3},
    urldate = {2020-04-04},
    journal = {Monthly Notices of the Royal Astronomical Society},
    author = {Mastrobuono-Battisti, Alessandra and Perets, Hagai B.},
    month = aug,
    year = {2017},
    note = {Publisher: Oxford Academic},
    pages = {3597--3609},
}

@article{agertz_fundamental_2007,
    title = {Fundamental differences between {SPH} and grid methods},
    volume = {380},
    issn = {0035-8711},
    url = {https://academic.oup.com/mnras/article/380/3/963/952655},
    doi = {10.1111/j.1365-2966.2007.12183.x},
    number = {3},
    urldate = {2017-11-17},
    journal = {Monthly Notices of the Royal Astronomical Society},
    author = {Agertz, Oscar and Moore, Ben and Stadel, Joachim and Potter, Doug and Miniati, Francesco and Read, Justin and Mayer, Lucio and Gawryszczak, Artur and Kravtsov, Andrey and Nordlund, Ake and Pearce, Frazer and Quilis, Vicent and Rudd, Douglas and Springel, Volker and Stone, James and Tasker, Elizabeth and Teyssier, Romain and Wadsley, James and Walder, Rolf},
    month = sep,
    year = {2007},
    pages = {963--978},
}

@article{marinova_mega-impact_2008,
    title = {Mega-impact formation of the {Mars} hemispheric dichotomy},
    volume = {453},
    copyright = {2008 Nature Publishing Group},
    issn = {1476-4687},
    url = {https://www.nature.com/articles/nature07070},
    doi = {10.1038/nature07070},
    language = {en},
    number = {7199},
    urldate = {2019-08-19},
    journal = {Nature},
    author = {Marinova, Margarita M. and Aharonson, Oded and Asphaug, Erik},
    month = jun,
    year = {2008},
    pages = {1216--1219},
}

@article{kegerreis_consequences_2018,
    title = {Consequences of {Giant} {Impacts} on {Early} {Uranus} for {Rotation}, {Internal} {Structure}, {Debris}, and {Atmospheric} {Erosion}},
    volume = {861},
    issn = {0004-637X},
    url = {https://doi.org/10.3847%2F1538-4357%2Faac725},
    doi = {10.3847/1538-4357/aac725},
    language = {en},
    number = {1},
    urldate = {2019-08-19},
    journal = {The Astrophysical Journal},
    author = {Kegerreis, J. A. and Teodoro, L. F. A. and Eke, V. R. and Massey, R. J. and Catling, D. C. and Fryer, C. L. and Korycansky, D. G. and Warren, M. S. and Zahnle, K. J.},
    month = jul,
    year = {2018},
    pages = {52},
}

@article{kurosaki_exchange_2018,
    title = {The {Exchange} of {Mass} and {Angular} {Momentum} in the {Impact} {Event} of {Ice} {Giant} {Planets}: {Implications} for the {Origin} of {Uranus}},
    volume = {157},
    issn = {1538-3881},
    shorttitle = {The {Exchange} of {Mass} and {Angular} {Momentum} in the {Impact} {Event} of {Ice} {Giant} {Planets}},
    url = {https://doi.org/10.3847%2F1538-3881%2Faaf165},
    doi = {10.3847/1538-3881/aaf165},
    language = {en},
    number = {1},
    urldate = {2019-06-04},
    journal = {The Astronomical Journal},
    author = {Kurosaki, Kenji and Inutsuka, Shu-ichiro},
    month = dec,
    year = {2018},
    pages = {13},
}

@article{fang_moon-forming_2025,
    title = {The moon-forming impact as a constraint for the inner {Solar} system’s formation},
    volume = {537},
    issn = {1745-3925},
    url = {https://doi.org/10.1093/mnrasl/slae109},
    doi = {10.1093/mnrasl/slae109},
    number = {1},
    urldate = {2025-01-09},
    journal = {Monthly Notices of the Royal Astronomical Society: Letters},
    author = {Fang, Tong and Bi, Rongxi and Zhang, Hui and Zhou, You and Reinhardt, Christian and Deng, Hongping},
    month = feb,
    year = {2025},
    pages = {L14--L20},
}

@article{reinhardt_forming_2022,
    title = {Forming iron-rich planets with giant impacts},
    volume = {517},
    issn = {0035-8711},
    url = {https://doi.org/10.1093/mnras/stac1853},
    doi = {10.1093/mnras/stac1853},
    number = {3},
    urldate = {2023-01-28},
    journal = {Monthly Notices of the Royal Astronomical Society},
    author = {Reinhardt, Christian and Meier, Thomas and Stadel, Joachim G and Otegi, Jon F and Helled, Ravit},
    month = dec,
    year = {2022},
    pages = {3132--3143},
}

@article{denton_capture_2025,
    title = {Capture of an ancient {Charon} around {Pluto}},
    volume = {18},
    copyright = {2025 The Author(s), under exclusive licence to Springer Nature Limited},
    issn = {1752-0908},
    url = {https://www.nature.com/articles/s41561-024-01612-0},
    doi = {10.1038/s41561-024-01612-0},
    language = {en},
    number = {1},
    urldate = {2025-02-26},
    journal = {Nature Geoscience},
    author = {Denton, C. Adeene and Asphaug, Erik and Emsenhuber, Alexandre and Melikyan, Robert},
    month = jan,
    year = {2025},
    note = {Publisher: Nature Publishing Group},
    pages = {37--43},
}

@article{hauck_curious_2013,
    title = {The curious case of {Mercury}'s internal structure},
    volume = {118},
    copyright = {©2013. American Geophysical Union. All Rights Reserved.},
    issn = {2169-9100},
    url = {https://agupubs.onlinelibrary.wiley.com/doi/abs/10.1002/jgre.20091},
    doi = {10.1002/jgre.20091},
    language = {en},
    number = {6},
    urldate = {2020-04-04},
    journal = {Journal of Geophysical Research: Planets},
    author = {Hauck, Steven A. and Margot, Jean-Luc and Solomon, Sean C. and Phillips, Roger J. and Johnson, Catherine L. and Lemoine, Frank G. and Mazarico, Erwan and McCoy, Timothy J. and Padovan, Sebastiano and Peale, Stanton J. and Perry, Mark E. and Smith, David E. and Zuber, Maria T.},
    year = {2013},
    note = {\_eprint: https://agupubs.onlinelibrary.wiley.com/doi/pdf/10.1002/jgre.20091},
    pages = {1204--1220},
}

@article{leleu_peculiar_2018,
    title = {The peculiar shapes of {Saturn}’s small inner moons as evidence of mergers of similar-sized moonlets},
    volume = {2},
    copyright = {2018 The Author(s)},
    issn = {2397-3366},
    url = {https://www.nature.com/articles/s41550-018-0471-7},
    doi = {10.1038/s41550-018-0471-7},
    language = {en},
    number = {7},
    urldate = {2021-03-17},
    journal = {Nature Astronomy},
    author = {Leleu, A. and Jutzi, M. and Rubin, M.},
    month = jul,
    year = {2018},
    note = {Number: 7
Publisher: Nature Publishing Group},
    pages = {555--561},
}

@article{lundborg_strength-size_1967,
    title = {The strength-size relation of granite},
    volume = {4},
    issn = {0148-9062},
    url = {http://www.sciencedirect.com/science/article/pii/0148906267900113},
    doi = {10.1016/0148-9062(67)90011-3},
    number = {3},
    urldate = {2018-08-03},
    journal = {International Journal of Rock Mechanics and Mining Sciences \& Geomechanics Abstracts},
    author = {Lundborg, N.},
    month = jul,
    year = {1967},
    pages = {269--272},
}

@article{brundage_implementation_2013,
    series = {Proceedings of the 12th {Hypervelocity} {Impact} {Symposium}},
    title = {Implementation of {Tillotson} {Equation} of {State} for {Hypervelocity} {Impact} of {Metals}, {Geologic} {Materials}, and {Liquids}},
    volume = {58},
    issn = {1877-7058},
    url = {http://www.sciencedirect.com/science/article/pii/S1877705813009594},
    doi = {10.1016/j.proeng.2013.05.053},
    urldate = {2018-05-14},
    journal = {Procedia Engineering},
    author = {Brundage, Aaron L.},
    month = jan,
    year = {2013},
    pages = {461--470},
}

@misc{thompson_m-aneos_2019,
    title = {M-{ANEOS}},
    url = {https://zenodo.org/record/3525030},
    urldate = {2022-02-14},
    publisher = {Zenodo},
    author = {Thompson, S. L. and Lauson, H. S. and Melosh, H. J. and Collins, G. S. and Stewart, S. T.},
    month = nov,
    year = {2019},
    doi = {10.5281/zenodo.3525030},
}

@article{melosh_hydrocode_2007,
    title = {A hydrocode equation of state for {SiO2}},
    volume = {42},
    issn = {1086-9379},
    url = {http://adsabs.harvard.edu/abs/2007M%26PS...42.2079M},
    doi = {10.1111/j.1945-5100.2007.tb01009.x},
    urldate = {2020-04-04},
    journal = {Meteoritics and Planetary Science},
    author = {Melosh, H. J.},
    year = {2007},
    pages = {2079--2098},
}

@misc{tillotsonMetallicEquationsState1962,
  title = {Metallic {{Equations}} of {{State}} for {{Hypervelocity Impact}}},
  author = {Tillotson, J. H.},
  year = {1962},
  publisher = {General Atomics Report, Nr. GA-3216}
}

@article{meloshImpactCrateringGeologic1989,
  title = {Impact Cratering: {{A}} Geologic Process},
  shorttitle = {Impact Cratering},
  author = {Melosh, H. J.},
  year = {1989},
  journal = {Research supported by NASA. New York, Oxford University Press (Oxford Monographs on Geology and Geophysics, No. 11), 1989, 253 p.},
  url = {http://adsabs.harvard.edu/abs/1989icgp.book.....M},
  urldate = {2021-04-03}
}

@article{kokuboEvolutionCircumterrestrialDisk2000,
  title = {Evolution of a {{Circumterrestrial Disk}} and {{Formation}} of a {{Single Moon}}},
  author = {Kokubo, Eiichiro and Ida, Shigeru and Makino, Junichiro},
  year = {2000},
  month = dec,
  journal = {Icarus},
  volume = {148},
  number = {2},
  pages = {419--436},
  issn = {0019-1035},
  doi = {10.1006/icar.2000.6496},
  url = {https://www.sciencedirect.com/science/article/pii/S0019103500964960},
  urldate = {2022-02-15},
  langid = {english}
}

@article{pahlevanEquilibrationAftermathLunarforming2007,
  title = {Equilibration in the Aftermath of the Lunar-Forming Giant Impact},
  author = {Pahlevan, Kaveh and Stevenson, David J.},
  year = {2007},
  month = oct,
  journal = {Earth and Planetary Science Letters},
  volume = {262},
  number = {3},
  pages = {438--449},
  issn = {0012-821X},
  doi = {10.1016/j.epsl.2007.07.055},
  url = {http://www.sciencedirect.com/science/article/pii/S0012821X07005006},
  urldate = {2020-07-03},
  langid = {english}
}

@article{kegerreisImmediateOriginMoon2022,
  title = {Immediate {{Origin}} of the {{Moon}} as a {{Post-impact Satellite}}},
  author = {Kegerreis, J. A. and {Ruiz-Bonilla}, S. and Eke, V. R. and Massey, R. J. and Sandnes, T. D. and Teodoro, L. F. A.},
  year = {2022},
  month = oct,
  journal = {The Astrophysical Journal Letters},
  volume = {937},
  number = {2},
  pages = {L40},
  publisher = {The American Astronomical Society},
  issn = {2041-8205},
  doi = {10.3847/2041-8213/ac8d96},
  url = {https://dx.doi.org/10.3847/2041-8213/ac8d96},
  urldate = {2023-02-16},
  langid = {english}
}

@article{jutzi_shape_2015,
    title = {The shape and structure of cometary nuclei as a result of low-velocity accretion},
    volume = {348},
    url = {https://www.science.org/doi/10.1126/science.aaa4747},
    doi = {10.1126/science.aaa4747},
    number = {6241},
    urldate = {2025-04-09},
    journal = {Science},
    author = {Jutzi, M. and Asphaug, E.},
    month = jun,
    year = {2015},
    note = {Publisher: American Association for the Advancement of Science},
    pages = {1355--1358},
}

@article{wimarsson_rapid_2024,
    title = {Rapid formation of binary asteroid systems post rotational failure: {A} recipe for making atypically shaped satellites},
    volume = {421},
    issn = {0019-1035},
    shorttitle = {Rapid formation of binary asteroid systems post rotational failure},
    url = {https://www.sciencedirect.com/science/article/pii/S0019103524002835},
    doi = {10.1016/j.icarus.2024.116223},
    urldate = {2025-04-09},
    journal = {Icarus},
    author = {Wimarsson, John and Xiang, Zhen and Ferrari, Fabio and Jutzi, Martin and Madeira, Gustavo and Raducan, Sabina D. and Sánchez, Paul},
    month = oct,
    year = {2024},
    pages = {116223},
}

@article{jutzi_formation_2017,
    title = {Formation of bi-lobed shapes by sub-catastrophic collisions - {A} late origin of comet {67P}’s structure},
    volume = {597},
    copyright = {© ESO, 2016},
    issn = {0004-6361, 1432-0746},
    url = {https://www.aanda.org/articles/aa/abs/2017/01/aa28964-16/aa28964-16.html},
    doi = {10.1051/0004-6361/201628964},
    language = {en},
    urldate = {2025-04-09},
    journal = {Astronomy \& Astrophysics},
    author = {Jutzi, M. and Benz, W.},
    month = jan,
    year = {2017},
    note = {Publisher: EDP Sciences},
    pages = {A62},
}

@article{jutzi_shape_2019,
    title = {The shape and structure of small asteroids as a result of sub-catastrophic collisions},
    volume = {177},
    issn = {0032-0633},
    url = {https://www.sciencedirect.com/science/article/pii/S0032063319300868},
    doi = {10.1016/j.pss.2019.07.009},
    urldate = {2025-04-09},
    journal = {Planetary and Space Science},
    author = {Jutzi, Martin},
    month = nov,
    year = {2019},
    pages = {104695},
}

@incollection{pierazzo_brief_2004,
    address = {Berlin, Heidelberg},
    title = {A {Brief} {Introduction} to {Hydrocode} {Modeling} of {Impact} {Cratering}},
    isbn = {978-3-662-06423-8},
    url = {https://doi.org/10.1007/978-3-662-06423-8_16},
    language = {en},
    urldate = {2025-04-10},
    booktitle = {Cratering in {Marine} {Environments} and on {Ice}},
    publisher = {Springer},
    author = {Pierazzo, Elisabetta and Collins, Gareth},
    editor = {Dypvik, Henning and Burchell, Mark J. and Claeys, Philippe},
    year = {2004},
    doi = {10.1007/978-3-662-06423-8_16},
    pages = {323--340},
}

@article{wissing_new_2020,
    title = {A new equation of state applied to planetary impacts - {I}. {Models} of planetary interiors},
    volume = {635},
    copyright = {© ESO 2020},
    issn = {0004-6361, 1432-0746},
    url = {https://www.aanda.org/articles/aa/abs/2020/03/aa35814-19/aa35814-19.html},
    doi = {10.1051/0004-6361/201935814},
    language = {en},
    urldate = {2020-04-04},
    journal = {Astronomy \& Astrophysics},
    author = {Wissing, Robert and Hobbs, David},
    month = mar,
    year = {2020},
    note = {Publisher: EDP Sciences},
    pages = {A21},
}

@article{golabekOriginMartianDichotomy2011,
  title = {Origin of the Martian Dichotomy and {{Tharsis}} from a Giant Impact Causing Massive Magmatism},
  author = {Golabek, Gregor J. and Keller, Tobias and Gerya, Taras V. and Zhu, Guizhi and Tackley, Paul J. and Connolly, James A. D.},
  year = {2011},
  month = sep,
  journal = {Icarus},
  volume = {215},
  number = {1},
  pages = {346--357},
  issn = {0019-1035},
  doi = {10.1016/j.icarus.2011.06.012},
  url = {https://www.sciencedirect.com/science/article/pii/S0019103511002211},
  urldate = {2025-04-24}
}

@article{meierOriginJupitersFuzzy2025,
doi = {10.3847/1538-4357/addbe6},
url = {https://dx.doi.org/10.3847/1538-4357/addbe6},
year = {2025},
month = {jul},
publisher = {The American Astronomical Society},
volume = {988},
number = {1},
pages = {7},
author = {Meier, Thomas and Reinhardt, Christian and Shibata, Sho and Müller, Simon and Stadel, Joachim and Helled, Ravit},
title = {On the Origin of Jupiter’s Fuzzy Core: Constraints from N-body, Impact, and Evolution Simulations},
journal = {The Astrophysical Journal}
}

@misc{Bussmann2025,
      title={The possibility of a giant impact on Venus}, 
      author={Mirco Bussmann and Christian Reinhardt and Cedric Gillmann and Thomas Meier and Joachim Stadel and Paul Tackley and Ravit Helled},
      year={2025},
      eprint={2508.03239},
      archivePrefix={arXiv},
      primaryClass={astro-ph.EP},
      url={https://arxiv.org/abs/2508.03239}, 
}

@article{Wunnemann2006,
title = {A strain-based porosity model for use in hydrocode simulations of impacts and implications for transient crater growth in porous targets},
journal = {Icarus},
volume = {180},
number = {2},
pages = {514-527},
year = {2006},
issn = {0019-1035},
doi = {https://doi.org/10.1016/j.icarus.2005.10.013},
url = {https://www.sciencedirect.com/science/article/pii/S0019103505004124},
author = {K. Wünnemann and G.S. Collins and H.J. Melosh},
}

@ARTICLE{Helled2025,
       author = {{Helled}, Ravit},
        title = "{Ice Giants}",
      journal = {arXiv e-prints},
         year = 2025,
        month = apr,
          eid = {arXiv:2504.18219},
        pages = {arXiv:2504.18219},
archivePrefix = {arXiv},
       eprint = {2504.18219},
 primaryClass = {astro-ph.EP},
       adsurl = {https://ui.adsabs.harvard.edu/abs/2025arXiv250418219H}
}

@article{gingold_smoothed_1977,
    title = {Smoothed particle hydrodynamics: theory and application to non-spherical stars},
    volume = {181},
    issn = {0035-8711},
    shorttitle = {Smoothed particle hydrodynamics},
    url = {https://academic.oup.com/mnras/article/181/3/375/988212},
    doi = {10.1093/mnras/181.3.375},
    language = {en},
    number = {3},
    urldate = {2020-04-23},
    journal = {Monthly Notices of the Royal Astronomical Society},
    author = {Gingold, R. A. and Monaghan, J. J.},
    month = dec,
    year = {1977},
    note = {Publisher: Oxford Academic},
    pages = {375--389},
}

@article{lucy_numerical_1977,
    title = {A numerical approach to the testing of the fission hypothesis},
    volume = {82},
    issn = {0004-6256},
    url = {http://adsabs.harvard.edu/abs/1977AJ.....82.1013L},
    doi = {10.1086/112164},
    urldate = {2020-04-23},
    journal = {The Astronomical Journal},
    author = {Lucy, L. B.},
    month = dec,
    year = {1977},
    pages = {1013--1024},
}

@incollection{rosenblatt_formation_2020,
    title = {The {Formation} of the {Martian} {Moons}},
    isbn = {978-0-19-064792-6},
    url = {https://oxfordre.com/planetaryscience/display/10.1093/acrefore/9780190647926.001.0001/acrefore-9780190647926-e-24},
    language = {en},
    urldate = {2025-04-09},
    booktitle = {Oxford {Research} {Encyclopedia} of {Planetary} {Science}},
    author = {Rosenblatt, Pascal and Hyodo, Ryuki and Pignatale, Francesco and Trinh, Antony and Charnoz, Sebastien and Dunseath, Kevin and Dunseath-Terao, Mariko and Genda, Hidenori},
    month = mar,
    year = {2020},
    doi = {10.1093/acrefore/9780190647926.013.24},
}

@article{canup_origin_2018,
    title = {Origin of {Phobos} and {Deimos} by the impact of a {Vesta}-to-{Ceres} sized body with {Mars}},
    volume = {4},
    url = {https://www.science.org/doi/10.1126/sciadv.aar6887},
    doi = {10.1126/sciadv.aar6887},
    number = {4},
    urldate = {2024-09-26},
    journal = {Science Advances},
    author = {Canup, Robin and Salmon, Julien},
    month = apr,
    year = {2018},
    note = {Publisher: American Association for the Advancement of Science},
    pages = {eaar6887},
}

@inproceedings{collins_improvements_2014,
    title = {Improvements to {ANEOS} for {Multiple} {Phase} {Transitions}},
    url = {https://ui.adsabs.harvard.edu/abs/2014LPI....45.2664C},
    urldate = {2025-09-10},
    author = {Collins, G. S. and Melosh, H. J.},
    month = mar,
    year = {2014},
    note = {ADS Bibcode: 2014LPI....45.2664C},
    pages = {2664},
}

@article{jutzi_forming_2011,
    title = {Forming the lunar farside highlands by accretion of a companion moon},
    volume = {476},
    copyright = {2011 Springer Nature Limited},
    issn = {1476-4687},
    url = {https://www.nature.com/articles/nature10289},
    doi = {10.1038/nature10289},
    language = {en},
    number = {7358},
    urldate = {2025-08-28},
    journal = {Nature},
    author = {Jutzi, M. and Asphaug, E.},
    month = aug,
    year = {2011},
    note = {Publisher: Nature Publishing Group},
    pages = {69--72},
}

@article{zhu_are_2019,
    title = {Are the {Moon}'s {Nearside}-{Farside} {Asymmetries} the {Result} of a {Giant} {Impact}?},
    volume = {124},
    copyright = {©2019. American Geophysical Union. All Rights Reserved.},
    issn = {2169-9100},
    url = {https://onlinelibrary.wiley.com/doi/abs/10.1029/2018JE005826},
    doi = {10.1029/2018JE005826},
    language = {en},
    number = {8},
    urldate = {2025-09-10},
    journal = {Journal of Geophysical Research: Planets},
    author = {Zhu, Meng-Hua and Wünnemann, Kai and Potter, Ross W.K. and Kleine, Thorsten and Morbidelli, Alessandro},
    year = {2019},
    note = {\_eprint: https://agupubs.onlinelibrary.wiley.com/doi/pdf/10.1029/2018JE005826},
    pages = {2117--2140},
}

@article{monaghan_smoothed_1992,
    title = {Smoothed particle hydrodynamics},
    volume = {30},
    issn = {0066-4146},
    url = {http://adsabs.harvard.edu/abs/1992ARA%26A..30..543M},
    doi = {10.1146/annurev.aa.30.090192.002551},
    urldate = {2017-03-15},
    journal = {Annual Review of Astronomy and Astrophysics},
    author = {Monaghan, J. J.},
    year = {1992},
    pages = {543--574},
}

@article{springel_smoothed_2010,
    title = {Smoothed {Particle} {Hydrodynamics} in {Astrophysics}},
    volume = {48},
    url = {http://dx.doi.org/10.1146/annurev-astro-081309-130914},
    doi = {10.1146/annurev-astro-081309-130914},
    number = {1},
    urldate = {2017-05-22},
    journal = {Annual Review of Astronomy and Astrophysics},
    author = {Springel, Volker},
    year = {2010},
    pages = {391--430},
}

@article{benz_origin_1986,
    title = {The origin of the moon and the single-impact hypothesis {I}},
    volume = {66},
    issn = {0019-1035},
    url = {http://www.sciencedirect.com/science/article/pii/0019103586900886},
    doi = {10.1016/0019-1035(86)90088-6},
    number = {3},
    urldate = {2018-05-14},
    journal = {Icarus},
    author = {Benz, W. and Slattery, W. L. and Cameron, A. G. W.},
    month = jun,
    year = {1986},
    pages = {515--535},
}

@article{canup_dynamics_2004,
    title = {Dynamics of {Lunar} {Formation}},
    volume = {42},
    url = {http://dx.doi.org/10.1146/annurev.astro.41.082201.113457},
    doi = {10.1146/annurev.astro.41.082201.113457},
    number = {1},
    urldate = {2017-05-09},
    journal = {Annual Review of Astronomy and Astrophysics},
    author = {Canup, Robin M.},
    year = {2004},
    pages = {441--475},
}

\end{document}